\documentclass[12pt]{article}

\usepackage{color}
\usepackage{amsmath}
\usepackage{amsfonts}
\usepackage{amssymb}
\usepackage{caption}
\usepackage{graphicx}
\usepackage{slashed}            % for slashed characters in math mode
\usepackage{subcaption}
\usepackage{xspace}				% For spacing after commands
\usepackage{cite}
\usepackage{bm}
\let\originalleft\left
\let\originalright\right
\renewcommand{\left}{\mathopen{}\mathclose\bgroup\originalleft}
\renewcommand{\right}{\aftergroup\egroup\originalright}

\usepackage[english]{babel}
\usepackage{fancyhdr}
\usepackage{amsmath}
\usepackage{amssymb}
\usepackage{amsfonts}
\usepackage{psfrag}
\usepackage[applemac]{inputenc}
\usepackage{graphicx}
\usepackage{xcolor}
\newcommand{\arXivold}[1]{\href{http://arxiv.org/pdf/#1}{{\tt #1}}}

\renewcommand{\tilde}{\widetilde} % dinky tildes look silly

\newcommand{\beq}{\begin{eqnarray}}
\newcommand{\eeq}{\end{eqnarray}}

\newcommand{\bag}{\begin{align}}
\newcommand{\eag}{\end{align}}

\usepackage[hypertexnames=false]{hyperref}		% kills links in index (if exists)

\begin{document}
\begin{titlepage}

\begin{center} %TITLE HERE
{\huge \bf  Cosmological Quasiparticles and the Cosmological Collider} 
\end{center}
\vspace*{0.4cm} 

\begin{center} % AUTHORS HERE
{\bf \   Jay Hubisz$^a$, Seung J. Lee$^{b,c}$, He Li$^a$, Bharath Sambasivam$^a$} 

% PLACES HERE
 {\it $^a$ Department of Physics, Syracuse University, Syracuse, NY  13244\\
       $^b$ Department of Physics, Korea University, Seoul 136-713, Korea \\
       $^c$  School of Physics, KIAS, Seoul 02455, Korea}

\vspace*{0.1cm}
% EMAILS HERE
{\tt  
 \href{mailto:jhubisz@syr.edu}{jhubisz@syr.edu},
 \href{mailto:s.jj.lee@gmail.com}{s.jj.lee@gmail.com},  
 \href{mailto:hli236@syr.edu}{hli236@syr.edu},
 \href{mailto:bsambasi@syr.edu}{bsambasi@syr.edu}}

\end{center}

\vglue 0.3truecm

\begin{abstract}
The interplay between cosmology and strongly coupled dynamics can yield transient spectral features that vanish at late times, but which may leave behind phenomenological signatures in the spectrum of primordial fluctuations.  Of particular interest are strongly coupled extensions of the standard model featuring approximate conformal invariance.  In flat space, the spectral density for a scalar operator in a conformal field theory is characterized by a continuum with scaling law governed by the dimension of the operator, and is otherwise featureless.  AdS/CFT arguments suggest that for large $N$, in an inflationary background with Hubble rate $H$, this continuum is gapped.  We demonstrate that there can be additional peak structures that become sharp and particle-like at phenomenologically interesting regions in parameter space, and we estimate their contribution to cosmological observables.  We find phenomena that are potentially observable in future experiments that are unique to these models, including displaced oscillatory features in the squeezed limit of the bi-spectrum.  These particles can be either fundamental, and localized to a UV brane, or composite at the Hubble scale, $H$, and bound to a horizon in the bulk of the 5D geometry.   We comment on how stabilization of conformal symmetry breaking vacua can be correlated with these spectral features and their phenomenology.
\end{abstract}

\end{titlepage}

\setcounter{equation}{0}
\setcounter{footnote}{0}
\setcounter{section}{0}

%%%%%%%%%%%%%%%%%%%%%%%%%%%%%%%%%%%%%%%%%%%%%%%%%%%%
%%%%%%%%%%%%%%%%%%%%%%%%%%%%%%%%%%%%%%%%%%%%%%%%%%%%
%%%%%%%%%%%%%%%%%%%%%%%%%%%%%%%%%%%%%%%%%%%%%%%%%%%%

\section{Introduction}
Strong dynamics and approximate conformal symmetry may play a crucial role in the generation of large hierarchies in fundamental physics~\cite{Randall:1999ee, Rattazzi:2000hs, Arkani-Hamed:2000ijo}. The AdS/CFT duality~\cite{Maldacena:1997re, Witten:1998qj, Gubser:1998bc} may in turn give perturbative control over strong dynamics in theories with a large number of gauge colors, where a large $N$ expansion corresponds to weakly coupled physics in an extra dimensional theory.  Randall-Sundrum models~\cite{Randall:1999ee} are 5D theories that are dual to strongly coupled 4D CFT's, and are the foundation for many interesting extensions of the SM~\cite{Contino:2003ve, Bellazzini:2015cgj, Bunk:2017fic, Kumar:2018jxz, Csaki:2018kxb, Csaki:2021gfm, Csaki:2022lnq}.

So far, collider experiments have yet to discover any direct evidence that models of this sort are behind nature's resolution of various puzzles of the standard model, and it may be decades before new collider experiments with center of mass energy sufficient to probe them will begin operating.  Experimental progress in cosmology, however, is rapidly developing, and offers opportunities to probe significant regions of the parameter space of such models.

The primordial power spectrum, in particular, offers an opportunity to probe energy scales that may be forever out of reach of terrestrial experiments due to the extremely high energy scales relevant during an early universe inflationary epoch.  Non-gaussian features of the curvature perturbations carry information about interactions of the inflaton with itself and with other particles with mass scales near or below the Hubble rate during inflation (see, for example, ~\cite{Maldacena:2002vr, deRham:2007jif, Arkani-Hamed:2015bza, Chen:2009zp, Chen:2010xka, Chen:2016uwp, Kumar:2017ecc, Baumann:2018muz, McAneny:2019epy, Aoki:2023tjm}).  In this work, we are interested in the imprint of strongly coupled, approximately conformal dynamics on primordial density fluctuations.

\begin{figure}[h!]
\center{
\includegraphics[width=0.5\textwidth]{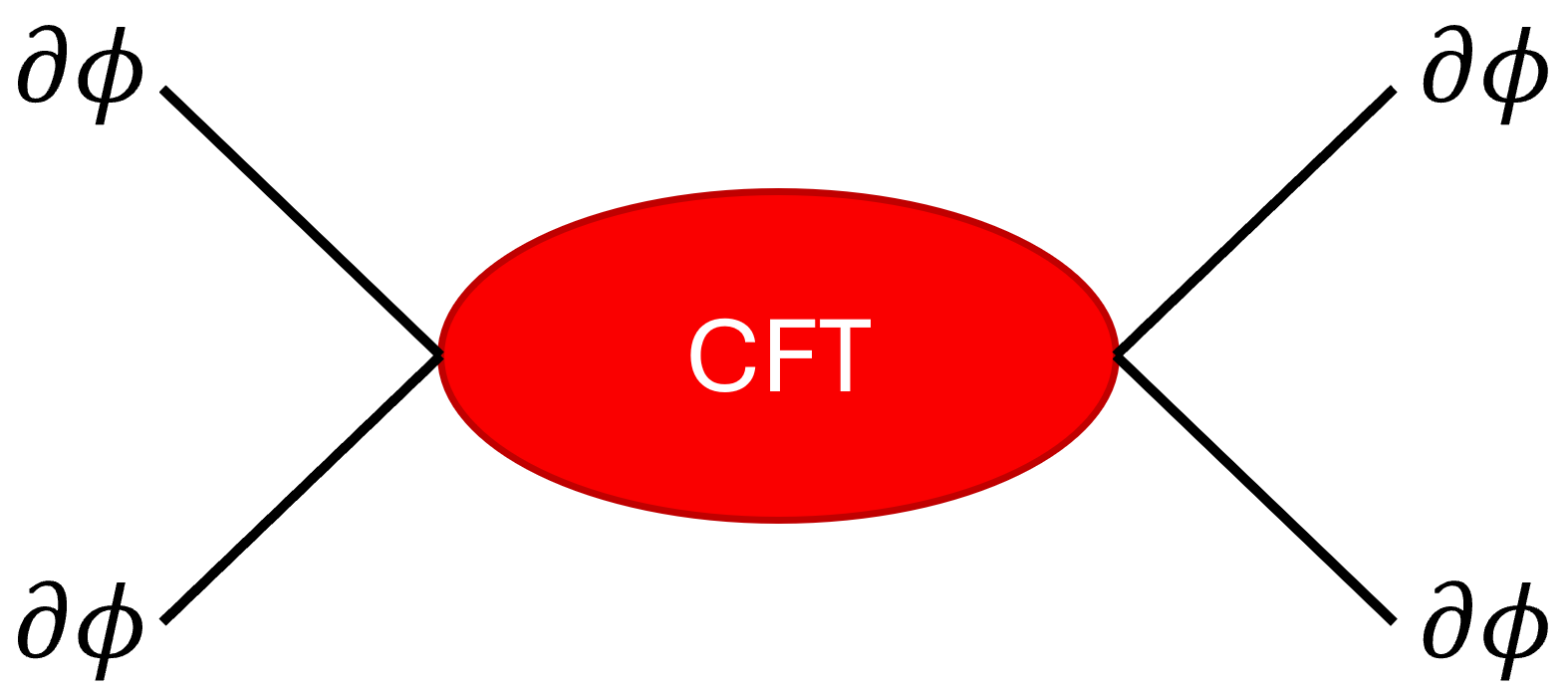}\hspace{0.5in}
\includegraphics[width=0.3\textwidth]{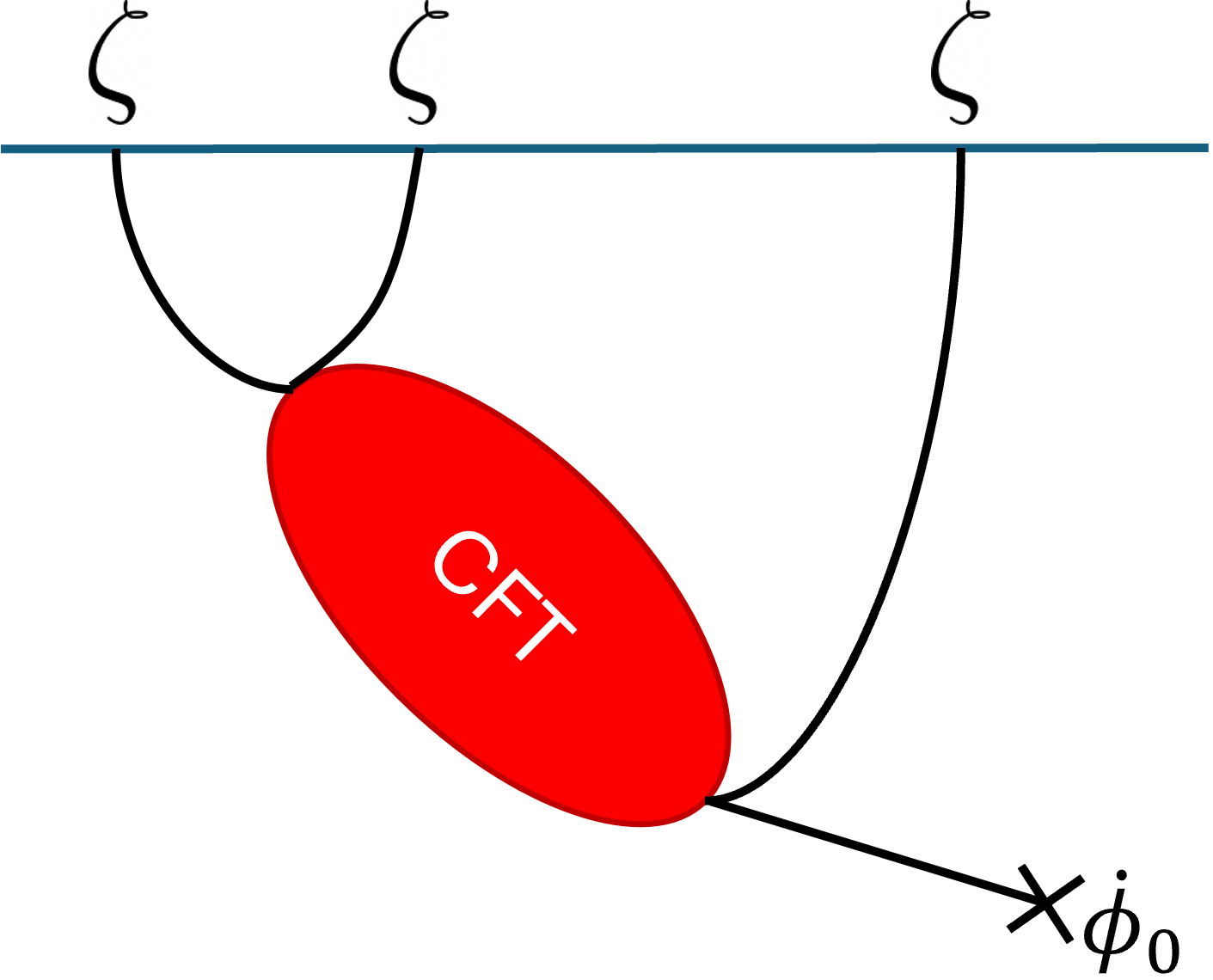}}
\caption{Diagrams corresponding to CFT modes being exchanged between inflaton particles.  In the second figure, one of the external inflatons is taken to be the time-dependent background of the rolling inflaton field, while the remaining 3 are the fluctuations about the background, and diagonalized to the scalar curvature perturbations.  The external lines are projected onto the time corresponding to the end of inflation.}
\label{fig:diagrams}
\end{figure}

Our universe is obviously not conformal at the energy scales we have thus far been able to probe, and therefore conformal symmetry must be somehow broken in the infrared (explicitly, spontaneously, or a combination of both).  However, if beyond the standard model (BSM) physics involves some sort of conformal sector, there may be an era in the early universe in which an approximate conformal symmetry is restored.  If, during the high energy epoch of inflation, conformal symmetry is unbroken (at least up to effects associated with the curvature of the universe itself), we might ask what sort of features would arise in the primordial spectrum of fluctuations due to interactions between the inflaton and this hypothetical conformal sector.  Such features could arise due to the exchange of CFT modes and the inflaton field, as shown in the cartoon in Fig.~\ref{fig:diagrams}.

To understand what sort of features in the power spectrum are created due to these interactions, we must calculate the two point function of operators associated with the CFT that the inflaton couples to.  The two point function of an operator in an \emph{unbroken} conformal field theory is determined by symmetry up to a single parameter, the scaling dimension of that operator.  In momentum space, the two point function of a scalar operator, $\phi$, in a 4D CFT is:
\beq
\langle \phi(x) \phi(x') \rangle \propto \int \frac{d^4p}{(2\pi)^4} e^{i p (x-x')} \frac{i}{(p^2+i\epsilon)^{2- \Delta}},
\eeq
and reduces to the propagator for the massless free scalar field when $\Delta$ saturates the unitarity bound, $\Delta \ge 1$.  The two-point function can be expressed in K\"all\'en-Lehmann form~\cite{Loparco:2023rug} as an integral over a spectral density, $\rho(\mu^2)$:
\beq
\langle \phi(x) \phi(x') \rangle = i\int  \frac{d^4p}{(2\pi)^4} e^{i p (x-x')} \int_0^\infty d(\mu^2) \frac{\rho(\mu^2)}{(p^2-\mu^2+i \epsilon)}, \text{~~~~~with~~~~~} \rho(\mu^2) = \frac{C(\Delta) }{(\mu^2)^{2-\Delta}}.
\eeq
The normalization factor $C$ vanishes as $\Delta \rightarrow 1$, so as to reproduce the spectral density for a massless free scalar particle, $\rho(\mu^2) = \delta(\mu^2)$. For $\Delta >1$, there is no particle excitation, and instead there is a continuum beginning at $\mu^2 = 0$, characterized by power-law scaling.

We expect that putting such a quantum field theory on a curved background must deform this spectral density to some degree, since the curvature of spacetime breaks conformal symmetry explicitly.  In fact, if we embed this theory in our cosmological universe, gravitational effects break conformal symmetry both in the UV and IR.  Conformal symmetry is broken explicitly near the Planck scale, when gravity becomes strong, and it is also broken in the IR by the finite Hubble rate associated with a non-vanishing stress-energy that sources cosmological solutions of the 4D Einstein equations.

\begin{figure}[h!]
\center{
\includegraphics[width=0.45\textwidth]{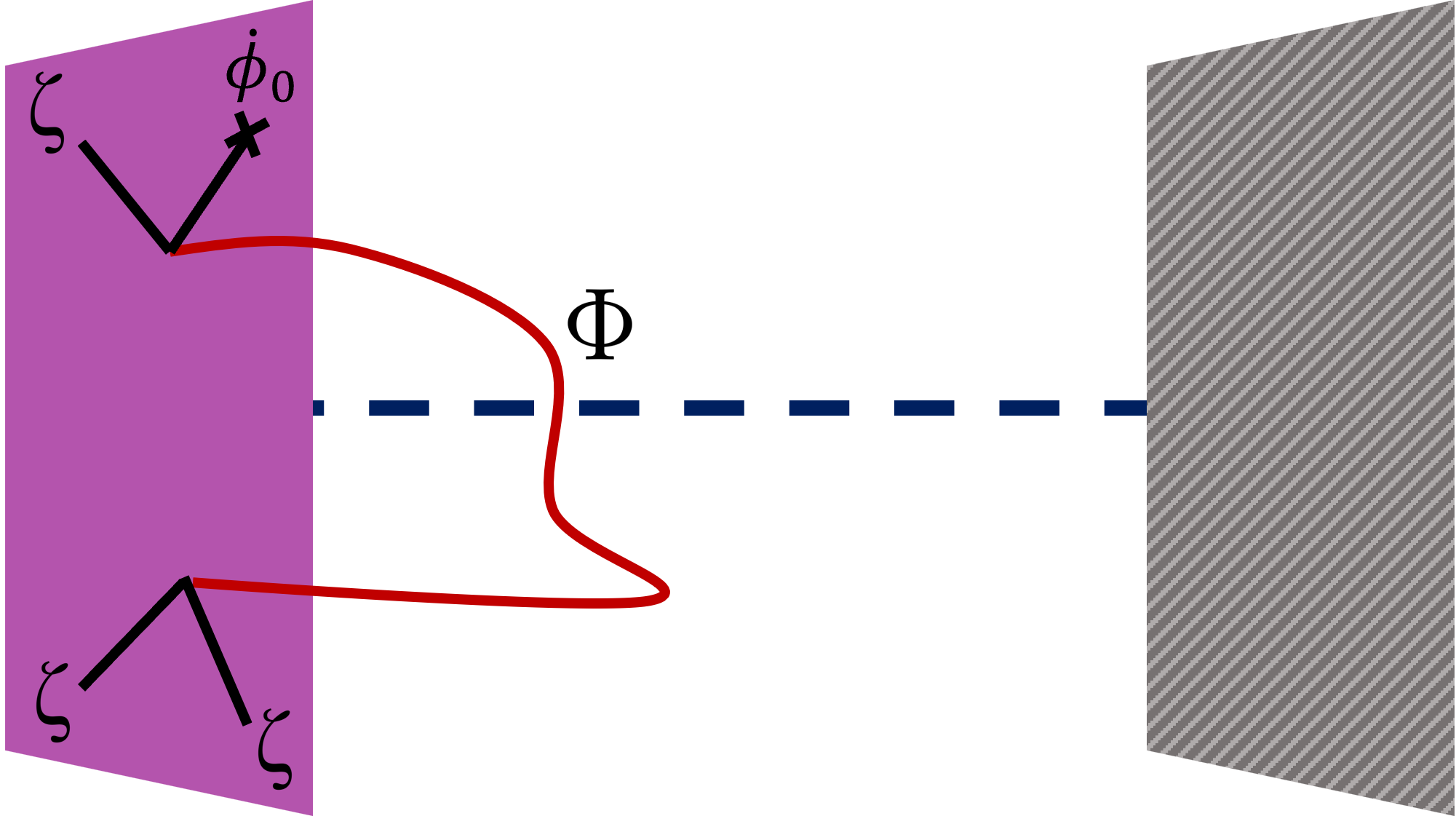}}
\caption{5D picture of the 3-point function of the UV-brane localized inflaton via the exchange of a bulk scalar field, $\Phi$. The UV brane is shown in purple. The horizon is shown in gray at a finite proper distance from the UV brane.}
\label{fig:5Ddiagram}
\end{figure}

While it is unclear how best to perform this calculation in the context of some unspecified strongly coupled 4D CFT coupled weakly to a fundamental sector, RS models offer an opportunity to perform the entire calculation in the context of a classical 5D theory, where UV brane-to-UV brane correlators correspond to the two-point function of some dual strongly coupled 4D CFT weakly coupled to 4D gravity.  In this case, the classical correlators in the 5D geometry are dual to taking into account conformal symmetry breaking effects, both in the UV and in the IR, at the quantum level.  In Figure~\ref{fig:5Ddiagram}, we show the diagram in 5D that corresponds to the 4D picture shown in Figure~\ref{fig:diagrams}.

On the 5D side of the duality, the IR 4D quantum effects are encapsulated in classical effects that deform the bulk geometry away from pure AdS in the ``infrared'' region of the geometry, deep in the bulk.  The geometry, in turn, changes the correlation functions, which encode the spectrum.  At strong coupling, non-perturbative effects due to cosmological expansion can be significant at scales close to the Hubble rate, and the spectrum can be quite different in comparison with the late-time description of the theory once inflation has ended and the universe has cooled.

In this work, we explore the phenomenology of inflationary dynamics of a 5D Randall-Sundrum theory with a bulk scalar field, such as one that might be responsible for the stabilization of the geometry after the epoch of inflation~\cite{Goldberger:1999un, Goldberger:1999uk}. In the case of a quasi-de-Sitter phase, under the presumption that there is no IR brane (or equivalently that the conformal symmetry is not spontaneously broken during inflation, and the strongly coupled theory is in a deconfined phase), the spectrum for the scalar field contains a gapped continuum~\cite{Kobayashi:2000yh, Riotto:2002yw, Koyama:2007as, Agashe:2020lfz, Gabadadze:2021dnk}. The Hubble rate, which breaks conformal symmetry explicitly, sets this gap.  

We demonstrate that there can additionally be a single massless particle below the gap, or a narrow peak in the spectral density if the mass of this particle overlaps with the continuum.  
This particle can be either UV or IR localized, depending on whether the bulk mass term is above or below $-3$ in units of the AdS curvature.  In the latter case, the particle is localized close to the horizon, and is a 5D analog of ``gravitational atom" solutions in 4D Schwarzchild geometries~\cite{Chia:2022udn}.

\begin{figure}[h!]
\center{
\includegraphics[width=0.85\textwidth]{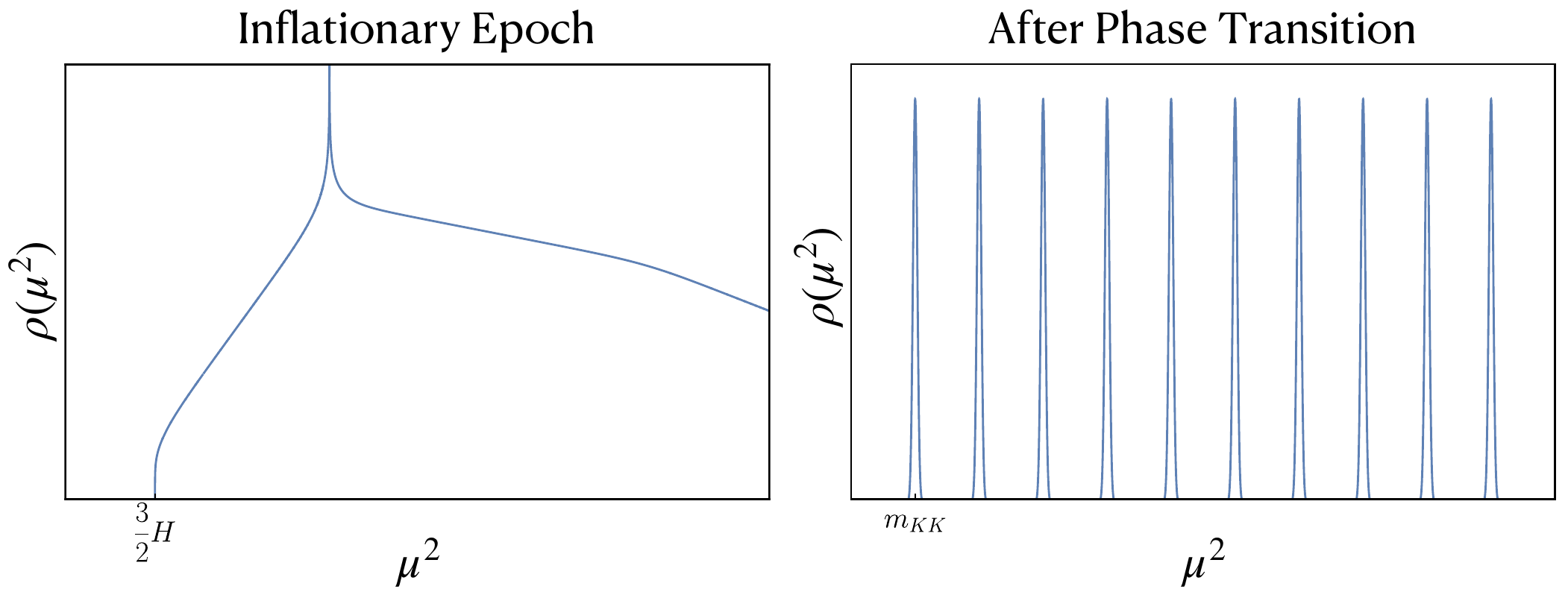}}
\caption{Here we show a cartoon of the spectral density corresponding to the brane-to-brane propagator for a 5D scalar field.  On the left is the picture during inflation, with Hubble rate $H$, which we presume corresponds to an epoch where the RS phase transition has not yet occurred.  The spectral density is that of a gapped continuum, with other features dependent on the model parameters.  On the right is the spectral density after the phase transition, when the spectrum is that of discrete Kaluza-Klein modes. }
\label{fig:rhocartoon}
\end{figure}

This is to be contrasted with the spectrum at late times, once the universe has cooled below the conformal phase transition temperature.  This spectrum might be a tower of resonances - the Kaluza-Klein modes of the 5D field (see Figure~\ref{fig:rhocartoon}), which are dual to a tower of composite particles near the confinement scale.  We study the phenomenological implications of such quasi-particles and their associated continuum on inflationary perturbations. The imprint of such physics on the bi-spectrum may eventually constrain the presence of large-N strong conformal dynamics during early universe cosmology.

This paper is organized as follows:  In Section~\ref{sec:infmodel}, we describe the basic setup, where inflation is driven by a fundamental UV brane localized scalar field.  In Section~\ref{sec:scalarsol}, we describe the spectrum of a 5D scalar field in the background of the inflationary cosmology.  In Section~\ref{sec:corrfun}, the correlation functions are given, along with the corresponding spectral density.    In Section~\ref{sec:coscoll}, we describe the effects of the strongly coupled conformal dynamics on non-gaussianities in the power spectrum of fluctuations arising from inflation.  In Section~\ref{sec:rad_stab}, we focus on models in which a stabilizing scalar field contributes to cosmological observables, and then we conclude.

%%%%%%%%%%%%%%%%%%%%%%%%%%%%%%%%%%%%%%%%%%%%%%%%%%%%
%%%%%%%%%%%%%%%%%%%%%%%%%%%%%%%%%%%%%%%%%%%%%%%%%%%%
%%%%%%%%%%%%%%%%%%%%%%%%%%%%%%%%%%%%%%%%%%%%%%%%%%%%

\section{Modeling Inflation}
\label{sec:infmodel}

Our model for inflation is taken to be as simple as possible, so that the dynamics we wish to describe are not conflated with novel inflationary dynamics.  We consider the physics of a fundamental inflationary sector that is coupled weakly to a strongly coupled CFT.  Schematically, our model can thus be described by the following lagrangian:
\beq
{\mathcal L} = {\mathcal L}_{\rm inf} + {\mathcal L}_{\rm CFT} + \sum_{ij} g_{ij} {\mathcal O}_{\rm inf}^i {\mathcal O}_{\rm CFT}^j,
\eeq
where ${\mathcal O}_{\rm inf}^i$ and ${\mathcal O}_{\rm CFT}^j$ are operators from the fundamental inflationary and strongly coupled CFT sectors, respectively.  This is the 4D theory as described at a high scale, which we take to be the Planck scale in our study.

In the 5D picture, the fundamental inflationary sector corresponds to a UV brane localized scalar field that drives slow-roll inflation via a slowly evolving mistune of the UV brane tension against a bulk cosmological constant.   The inflationary action is therfore a 5D Einstein-Hilbert action on a space with one brane, and a scalar field action on that brane:
\beq
S= -\int d^5x \sqrt{g} \left[ \Lambda + \frac{1}{2 \kappa^2} R \right] + \int d^4x \sqrt{g_0} \left[ \frac{1}{2} (\partial \varphi)^2 - \lambda(\varphi) \right]
\eeq
We take the 5D cosmological constant to be parameterized as $\Lambda = -\frac{6 k^2}{\kappa^2}$, where $k$ is the curvature of AdS, $\kappa$ is the 5D gravitational constant, and $g_0$ is the induced metric on the UV brane.

The equation of motion for the boundary localized scalar field, under the ansatz of a general 5D time dependent metric is
\beq
\ddot{\varphi} + 3  H \dot{\varphi} + \frac{\partial \lambda(\varphi)}{\partial \varphi}  = 0,
\eeq
where $H$ is the Hubble rate associated with the cosmology on the UV brane, and the time derivatives are with respect to proper time on the brane.

The bulk equations of motion and the junction conditions for the brane combine to give an FLRW equation on the UV brane:
\beq
H^2 + \frac{1}{2} \dot{H}= \frac{\kappa^4}{36} \lambda^2(\varphi) \left( 1- \frac{\dot{\varphi}^2}{\lambda(\varphi)} \right) \left( 1+ \frac{\dot{\varphi}^2}{2 \lambda(\varphi)} \right) + \frac{\kappa^2}{6} \Lambda.
\eeq
In the case of slow-roll, a good approximation to the dynamics is that of \emph{static} mistuned RS-II, where there is only a single UV brane, and the 4D slices of this geometry are de-Sitter~\cite{Himemoto:2000nd, Koyama:2007as}.  The Hubble rate at a given field value $\varphi$ is thus given by the mistune:
\beq
H^2 \approx \frac{\kappa^4}{36} \lambda^2(\varphi) - k^2.
\eeq

For constant mistune, an ansatz for the 5D metric is thus
\beq
ds^2 = \frac{1}{(kz)^2}\left(  dt^2 -e^{2 H t} d\vec{x}^2 - \frac{dz^2}{G^2(z)} \right)
\label{eq:metansatz}
\eeq
with $H$ as the inflationary Hubble rate at some instant in time during inflation.  $G(z)$ is a function encoding deviations of the geometry from pure AdS space due to the finite Hubble rate.  The limit of pure AdS with curvature $k$ is obtained by taking $G = 1$, corresponding to $H=0$.  The zz-Einstein equation for the case of non-zero $H$ yields a simple expression for $G(z)$:
\beq
G(z) = \sqrt{1+ H^2 z^2}.
\eeq
This is a geometry of finite size.  With the UV brane localized at $z = k^{-1}$, the metric has a singularity at $z \rightarrow \infty$, corresponding to a horizon, and the length of the extra dimension is
\beq
L = \int_{1/k}^{\infty} \frac{1}{k z G} dz = k^{-1} \sinh^{-1} \frac{k}{H} \approx k^{-1} \log \frac{2 k}{H}.
\eeq
The finite size of the observable universe, $H^{-1}$, acts as an infrared cutoff for the geometry.  We therefore expect that the spectrum in the theory will be gapped.

In the next section, we calculate details of the spectrum of a scalar field propagating in this geometry.

%%%%%%%%%%%%%%%%%%%%%%%%%%%%%%%%%%%%%%%%%%%%%%%%%%%%
%%%%%%%%%%%%%%%%%%%%%%%%%%%%%%%%%%%%%%%%%%%%%%%%%%%%
%%%%%%%%%%%%%%%%%%%%%%%%%%%%%%%%%%%%%%%%%%%%%%%%%%%%

\section{Scalars in AdS-dS}
\label{sec:scalarsol}

We now consider a 5D bulk scalar field in the inflating geometry described in the previous section.  This is a field that is added in addition to the inflaton, and could, for example, be playing the role of geometry stabilization~\cite{Goldberger:1999un, Goldberger:1999uk} in the late-time spontaneously broken phase of the theory.  The spectrum always contains a continuum, and there may also be particles below the gap to this continuum.  We first demonstrate the existence of the continuum and then study the existence of particle states that may lie below this continuum.  The particles can be either UV localized or localized near the horizon, depending on the value of the 5D scalar mass term.
\begin{center}
{\bf The Gap}
\end{center}
As the 5D geometry is of finite size, we expect the theory to be gapped.  Here we show that the spectrum contains a continuum that begins at energy $\bar{\mu} = \frac{3}{2} H$.  To show this, it is convenient to switch to coordinates where the dependence on the extra dimensional coordinate is an overall conformal factor:
\beq
ds^2 = e^{-2 A(w)} \left[ dt^2 - e^{2 H t} -dw^2 \right],\text{ with } e^{-A(w)} = \frac{H}{k \sinh H w}.
\eeq

The bulk scalar equation of motion in the inflating geometry is then given by:
\beq
-\phi''+3 A' \phi' + m^2 e^{-2A(w)} \phi = -\Box_{dS_4} \phi \equiv \mu^2 \phi.
\eeq
We put this equation of motion into Schr\"odinger form by performing a rescaling of the field $\phi = \tilde{\phi} e^{\frac{3}{2} A(w)}$, in which case the equation of motion becomes:
\beq
-\tilde{\phi}'' + \left[ m^2 e^{-2A} +\frac{9}{4} A'^2 - \frac{3}{2} A'' \right] \tilde{\phi} = \mu^2 \tilde{\phi},
\eeq
For the dS brane geometry we study, this is given by
\beq
-\tilde{\phi}'' + H^2 \left[ \frac{9}{4} \coth^2 (H w) + \frac{3+2m^2}{2 \sinh^2(Hw)} \right] \tilde{\phi} = \mu^2 \tilde{\phi}
\eeq
The $w \rightarrow \infty$ asymptotics of the ``potential" term determine whether the KK spectrum is discrete or a continuum.  Since this potential asymptotes to $\frac{9}{4} H^2$ as $w \rightarrow \infty$, we know there is a gapped continuum beginning at $\bar{\mu}^2 = \frac{9}{4} H^2$~\cite{Falkowski:2008yr}.

We note briefly that this value for the gap is not completely robust.  For example, a 5D bulk coupling between the scalar field and the scalar curvature can be introduced:
\beq
{\mathcal L} \ni -\xi R \phi^2,
\eeq
where $\xi$ is a dimensionless constant.  This modifies the effective potential in the Schroedinger problem above, and the gap is shifted:
\beq
\bar{\mu}^2 = \left[ \frac{9}{4} - 12 \xi \right]H^2.
\eeq
The gap vanishes when $\xi = 3/16$.~\footnote{The ``conformal" value for the scalar curvature coupling in $d$ dimensions is $\xi_c = \frac{(d-2)}{4 (d-1)}$, corresponding to $3/16$ in 5D.} For $\xi > 3/16$, there would be an instability, corresponding to the existence of a tachyonic portion of the continuum, and we would need to determine the true vacuum of the theory.   Additionally, if the scalar mass is given a profile $m= m(w)$, perhaps through a coupling to another scalar field which obtains a $w$-dependent vev, then it can also potentially shift the value of $\bar{\mu}$.

For the rest of this work, we focus on the scenario with $\xi = 0$, and where the bulk mass term is constant.

\begin{center}
{\bf Quasiparticles:  UV Localized}
\end{center}

Additional details of the spectrum are determined by solving the scalar equation of motion subject to the boundary conditions.  Near the horizon in the IR region, the boundary condition is normalizability of the solutions.  Above the gap, $\bar{\mu} = \frac{3}{2} H$, all solutions are normalizable (in the Fourier sense), explaining the existence of a continuum. 

It is possible for there to be a light mode that lies below the gap.  It is clear from the Schr\"odinger formulation of the eigenvalue problem that there will \emph{always} be an exponentially decaying and normalizable solution in the regime where the potential approaches a finite constant, $\frac{9}{4} H^2$.  The UV boundary conditions then determine whether there are any eigenvalues consistent with masses in this regime.  A very interesting aspect of this is that such solutions can be present even when the mode is \emph{not} normalizable in the $H \rightarrow 0$ limit.  That is, there can be particles that are there due to the expansion of the universe.  We will discuss this possibility in the next subsection.

To look for the presence of particles, we search for solutions to the equation of motion corresponding to eigenvalues below the gap, $\frac{9}{4} H^2$.  To study these solutions to the equation of motion, it is beneficial to work with a field that has no time dependance in its normalization, and where the metric is in a slightly different form.  Switching to conformal time, the metric is:
\beq
ds^2 = \frac{1}{(kz)^2} \left[a^2(\eta) dx_4^2 - \frac{dz^2}{G^2(z)} \right].
\eeq
 where $a(\eta) = -\frac{1}{H \eta}$.  Then, in terms of a rescaled field $v(t,y) = a \phi$, the equation of motion separates as:
\beq
\frac{1}{a^2}\left[ \ddot{v} + \left( k^2  - \frac{\ddot{a}}{a} \right) v \right]  - G^2 \left[ v'' - \left( \frac{3}{z}- \frac{G'}{G} \right) v' - \frac{m^2}{(kz)^2 G^2} v \right] = 0.
\eeq

We separate variables: $v = \sigma(\eta) \varphi(y)$.  In this case, we satisfy
\begin{align}
& \varphi''- \left(\frac{3}{z}- \frac{H^2 z^2}{1+ H^2 z^2} \right) \varphi' - \frac{1}{H^2 z^2} \frac{\left( m^2 - z^2 \mu^2 \right)}{1+ H^2 z^2} \varphi=0 \nonumber \\
& \ddot{\sigma} + \left( k^2 - \frac{2- \left(\frac{\mu}{H}\right)^2}{\eta^2} \right)\sigma = 0
\end{align}

The time dependent piece, $\sigma$, is solved by Hankel functions:
\beq
\sigma = \sqrt{\eta} H^{(1,2)}_{\tilde{\gamma}} (| k \eta |),
\eeq
where, for convenience, we have defined $\tilde{\gamma} = \sqrt{\frac{9}{4}-\left(\frac{\mu}{H} \right)^2} = -i \gamma$.
Only the $H^{(1)}$ solution satisfies the Bunch-Davies initial condition, and we employ this assumption for the vacuum state of all of the modes.

The solutions for the spatial wave-function are given by hypergeometric functions:
\beq
\varphi_\pm = z^{2 \pm \nu} ~_2 F^1 \left[ \frac{1}{4} -  \frac{\tilde{\gamma}}{2} \pm \frac{\nu}{2},\frac{1}{4} + \frac{\tilde{\gamma}}{2} \pm \frac{\nu}{2},1 \pm \nu,-H^2 z^2 \right]
\eeq
where $\nu^2 = 4+m^2$.  A given solution for certain boundary conditions is given by
\beq\label{eq:ScalarSol}
\varphi = N \left( \varphi_+ + \alpha \varphi_- \right)
\eeq
where $N$ is a normalization factor, and $\alpha$ is a $\tilde{\gamma}$-dependent coefficient to be determined by the boundary conditions.  Normalizability and the UV brane boundary condition together determine both the mass of the particle and the value of $\alpha$.

We first discuss the case where $\nu > 1$.  In this case, any particle solution is UV localized, so we can study the equation of motion in the limit $z H \ll 1$.    In this regime, the potential and boundary condition are quite simple, and the effects of Hubble expansion are particularly easy to take into account.  They simply shift the bulk potential by a finite constant, and contribute as an effective shift in $m_0^2$ which modifies the boundary condition.  The result is:
\beq\label{eq:mu2approx}
\mu^2 = (\nu-1) \left( m_0^2 - 2 (2-\nu) \right) +2 (2-\nu) H^2 + {\mathcal O}(H^4),
\eeq
which is valid so long as the mode is UV localized.  This is the case so long as $\nu > 1$, and not very close to that threshold.  It should be noted that a rather high degree of fine tuning must be present to obtain a particle below the gap when $\nu> 1$.  From~\ref{eq:mu2approx}, we see that the mistune, $\delta \equiv m_0^2 - 2 (2-\nu)$, must be small in units of $H^2$, and experimental constraints require the Hubble rate during inflation to be small compared to the Planck scale.  If physics is supersymmetric at this scale, however, supersymmetry fixes the the mistune to $\delta = 0$~\cite{Gherghetta:2000qt, Marti:2001iw}. This, however, comes in addition to a constraint on a bulk scalar curvature term that sets the gap to zero~\cite{Cacciapaglia:2008bi}.

If $\mu^2$ is below the gap, this corresponds to an additional solution to the eigenvalue equation (and a bound state in the Schr\"odinger problem).  The first two terms in the expansion in $H$ are then both of the same order, meaning that the effects of Hubble expansion contribute to an ${\mathcal O}(1)$ shift in the mass of the particle.  

What is the fate of this particle if the boundary conditions are such that this mass would lie inside the continuum? In this case, there is no separate eigenvalue for this light particle.  Instead the particle ``lives in the continuum," and can appear as a sharp feature in the spectral density -- see Section~\ref{sec:corrfun} for details.  In the analysis of the Schr\"odinger problem, there is no bound state, but there \emph{is} a metastable state that decays into the continuum.

How do we interpret this?  In the language of the AdS/CFT correspondence, this bound state corresponds to the existence of a fundamental particle in the theory (the part of the field $\phi$ on the UV brane).  This bound state mixes with the CFT states, however, so the solutions to the equations of motion correspond to the diagonalized spectrum after taking this mixing into account.  This is strongly reminiscent of quasiparticles in condensed matter, where there are collective excitations of numerous fundamental degrees of freedom.  In this case the particle excitation is composed of a fundamental scalar particle that is ``dressed" with states from the CFT.  The CFT is sensitive to the IR deformation associated with the Hubble expansion, which can thus backreact onto the mass of the particle eigenstate.  If that particle becomes heavy (above the gap), then there is no separate eigenstate corresponding to the particle as it simply decays to the continuum.  There is still a particle-like behavior:  the state exhibits a finite width (already at the classical level in the 5D picture) due to its mixing with the modes of the CFT.

When the boundary mass is tuned so that $\delta \lesssim H^2$, the mass receives a sizable correction from Hubble expansion.   That there is a correction is no surprise:  any field should receive quantum corrections that are sensitive to the IR scales present, and that contribute to the effective action.  The correction here, however, is classical in the 5D picture. It is not suppressed by loop factors due to the fact that the 5D geometric solutions are describing the dual dynamics of a strongly coupled gauge theory through the AdS/CFT correspondence.  

Since the sign of the ${\mathcal O} (H^2)$ correction can be either positive or negative, the stability of the solution hinges on details of the interplay between the strong dynamics and the cosmology.  This means that significant events (like phase transitions) can occur in otherwise sequestered sectors due purely to the effects of cosmological back-reaction.  The sign of the correction term is negative if $\nu > 2$, and positive otherwise.  In the AdS/CFT picture, the scaling dimensions of the operator corresponding to the bulk scalar field is given by $\Delta_\pm = 2 \pm \nu$, where the sign depends on boundary conditions.  As $\nu$ passes upwards through 2, the $\Delta_+$ scaling dimension goes from being relevant to irrelevant, so the sign of the correction to the mass depends on the renormalization group properties of the operator dual to the bulk scalar field.

\begin{center}
{\bf Cosmological Quasiparticles: $\nu < 1$}
\end{center}

\begin{figure}[h!]
\begin{center}
\includegraphics[width=0.45\textwidth]{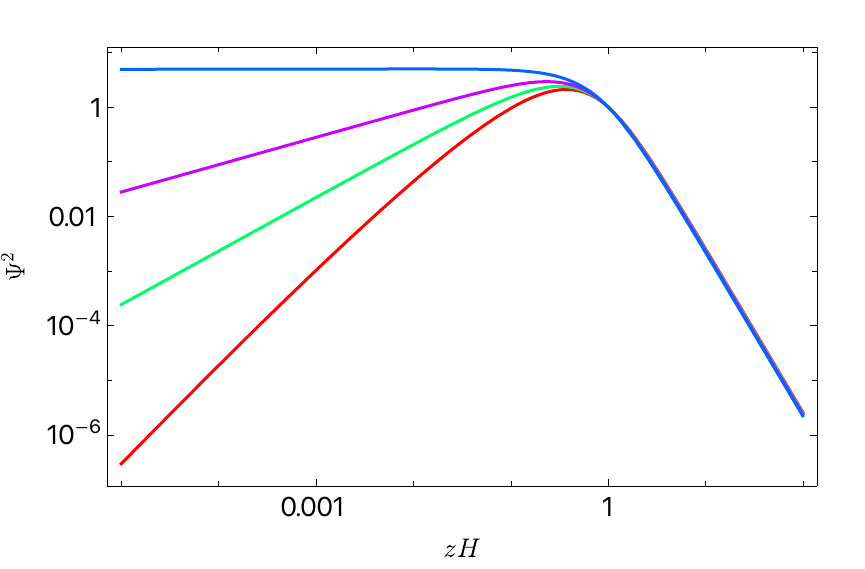}
\includegraphics[width=0.45\textwidth]{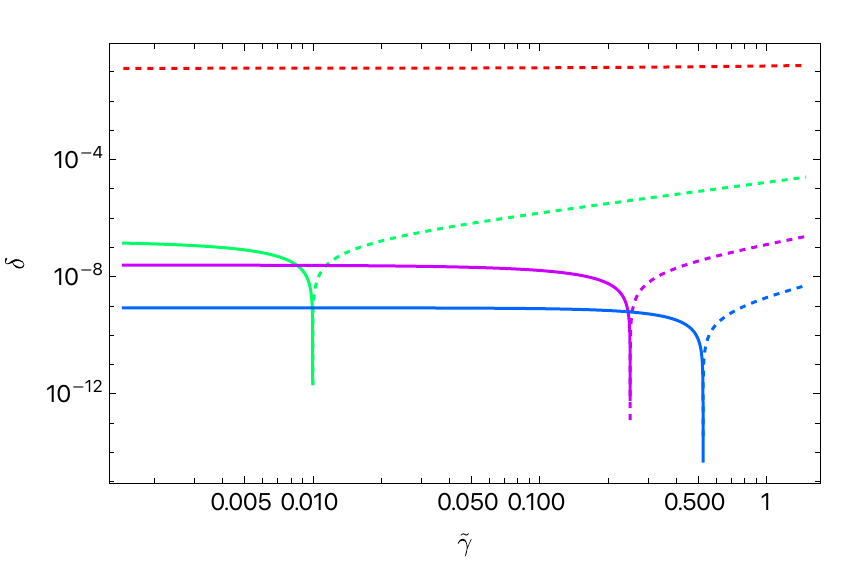}
\end{center}
\caption{We plot the squared wave functions (left) of modes with vanishing mass eigenvalue during inflation for various values of the $\nu$ parameter.  The wave functions are also multiplied by metric factors such that a decaying profile in the deep IR corresponds to normalizability.  The values of $\nu$ are $0~\text{(red)},1/2~\text{(green)},3/4~\text{(purple)},1~\text{(blue)}$.  The near horizon region corresponds to $zH \gtrsim 1$.  The plot on the right shows the required $\delta$ values to obtain mass $\mu$, with the horizontal axis reflecting the $\tilde{\gamma}= \sqrt{9/4-\mu^2/H^2}$ for a given $\mu$.  The Hubble rate is taken to be $10^{14}$~GeV.}
\label{fig:quasipWF}
\end{figure}

When $\nu <1$, there are no small-eigenvalue solutions to the equation of motion if the geometry is not truncated in the IR.  This is because the solutions in the near-AdS region of the space, $zH \ll 1$, scale like $z^{2-\nu}$.  For $\nu <1$, such modes are not normalizable if the geometry is not cut off at finite distance from the UV brane.  In the case of the inflating geometries we are analyzing, the presence of the horizon shuts off the bulk and permits the existence of IR-localized modes.   These modes are analogous to ``gravitational atom" solutions in Schwarzchild black-hole geometries for light scalar fields~\cite{Chia:2022udn}, since these modes are localized near the horizon in the bulk.  We plot the wavefunctions of these solutions for various values of the scalar bulk mass term in Figure~\ref{fig:quasipWF}, and we also display the relation between the mistune parameter, $\delta$ and the mass eigenvalue of the quasi-particle.  

As such particles have exponentially small overlap with the UV brane, they are to be interpreted in the CFT picture as predominantly composite modes of strongly coupled near-conformal dynamics.  Their compton wavelength is naturally of order the size of the universe, We thus dub them ``cosmological quasiparticles" as they appear to be transient in nature - a feature of the inflationary epoch that is to be eventually supplanted by the dynamics of conformal symmetry breaking that must occur to give rise to the standard model spectrum of particles at late times.

Since these quasi-particle modes are localized in the IR, we expect that the contribution to the masses of such particles from the UV brane mistune are suppressed.  The mass instead dominantly arises from the Hubble rate, and is expected to track it as it evolves with time.  We show the relation between the mistune parameter and the mass eigenvalue in the second plot in Figure~\ref{fig:quasipWF}.

%%%%%%%%%%%%%%%%%%%%%%%%%%%%%%%%%%%%%%%%%%%%%%%%%%%%
%%%%%%%%%%%%%%%%%%%%%%%%%%%%%%%%%%%%%%%%%%%%%%%%%%%%
%%%%%%%%%%%%%%%%%%%%%%%%%%%%%%%%%%%%%%%%%%%%%%%%%%%%

\section{Correlation functions and the Spectral Density}
\label{sec:corrfun}

The spectral density of a scalar operator in an unbroken CFT is determined by its scaling dimension, and is given by $\rho(\mu^2) = C(\Delta) (\mu^2)^{\Delta-2}$ in four dimensions, where $C(\Delta)$ is a normalization.  In the cosmological RS models we are considering, a richer phenomenology for the spectral density (as determined in our case by computation of the UV brane-to-brane propagator) is typically found.  The first reason is that the UV brane acts as a cutoff for the CFT, and it explicitly breaks the isometries of AdS.  The spectral density thus exhibits features corresponding to the breaking of conformal invariance.  Additionally, the cosmological Hubble rate breaks the CFT in the infrared.  The spectral density is thus empty (up to possible particle contributions) from $\mu^2 = 0$ up to the gap created by the Hubble expansion, $\bar{\mu} = \frac{3}{2}H$.  Particle or particle-like features also appear in the spectral density, and we will see that their presence and phenomenology can depend on both of these UV and IR cutoffs.

There are two regions of parameter space that we will explore separately.  The reason for this division arises from the holographic interpretation of the 5D model.  In the AdS-CFT correspondence, the boundary-to-boundary correlator for a scalar field in 5D exhibits one of two possible behaviors, depending on the UV boundary conditions.  The solutions to the scalar equation of motion at low momentum have two different scalings,
\beq
\Delta_\pm= 2 \pm \sqrt{4 + m^2}.
\label{eq:scalings}
\eeq
These $\Delta$'s are known to correspond to the scaling dimensions in a dual CFT, and the unitarity bound, $\Delta \ge 1$ implies that this interpretation is valid for both scalings when $-4 \le m^2 \le -3$.\footnote{The mass$^2$ term must be above $-4$ (the Breitenlohner-Freedman bound~\cite{Breitenlohner:1982bm, Breitenlohner:1982jf}) in order to have solutions that exhibit unitary behavior.}  The region $m^2 > -3$ has a different behavior that we will discuss momentarily.  The two scalings, $\Delta_\pm$ exist because the 5D model with a single scalar field with $-4 \le m^2 \le -3$ actually describes \emph{two} CFT's, with each of them associated with a different choice of boundary action for the scalar field~\cite{Klebanov:1999tb, Mueck:1999kk, Kaplan:2009kr}. This is in the case of an AdS space with the UV brane taken to the AdS boundary.  

In our case, the UV brane truncates the geometry, and can encode explicit breaking of the CFT at a specific high scale associated with the position of the brane.
 The UV-brane boundary condition for the scalar field then can be interpreted as a $\beta$-function that serves to create an RG flow between UV and IR CFT's where the scaling dimension of the operator flows from $\Delta_-$ in the UV, to $\Delta_+$ in the IR.  The $\Delta_-$ scaling then corresponds to the dimension of an operator at an unstable UV fixed point, with the $\Delta_+$ scaling reflecting the dimension at the IR fixed point in the flow.  These boundary conditions are determined by the UV potential for the bulk scalar field.  If, as in the section above, we consider a simple mass term, then the mistune in the brane mass term against the bulk mass, $\delta m_0^2 = m_0^2 - 2 (2- \nu)$, can be interpreted as this $\beta$ function.

\begin{figure}[h!]
\center{
\includegraphics[width=0.45\textwidth]{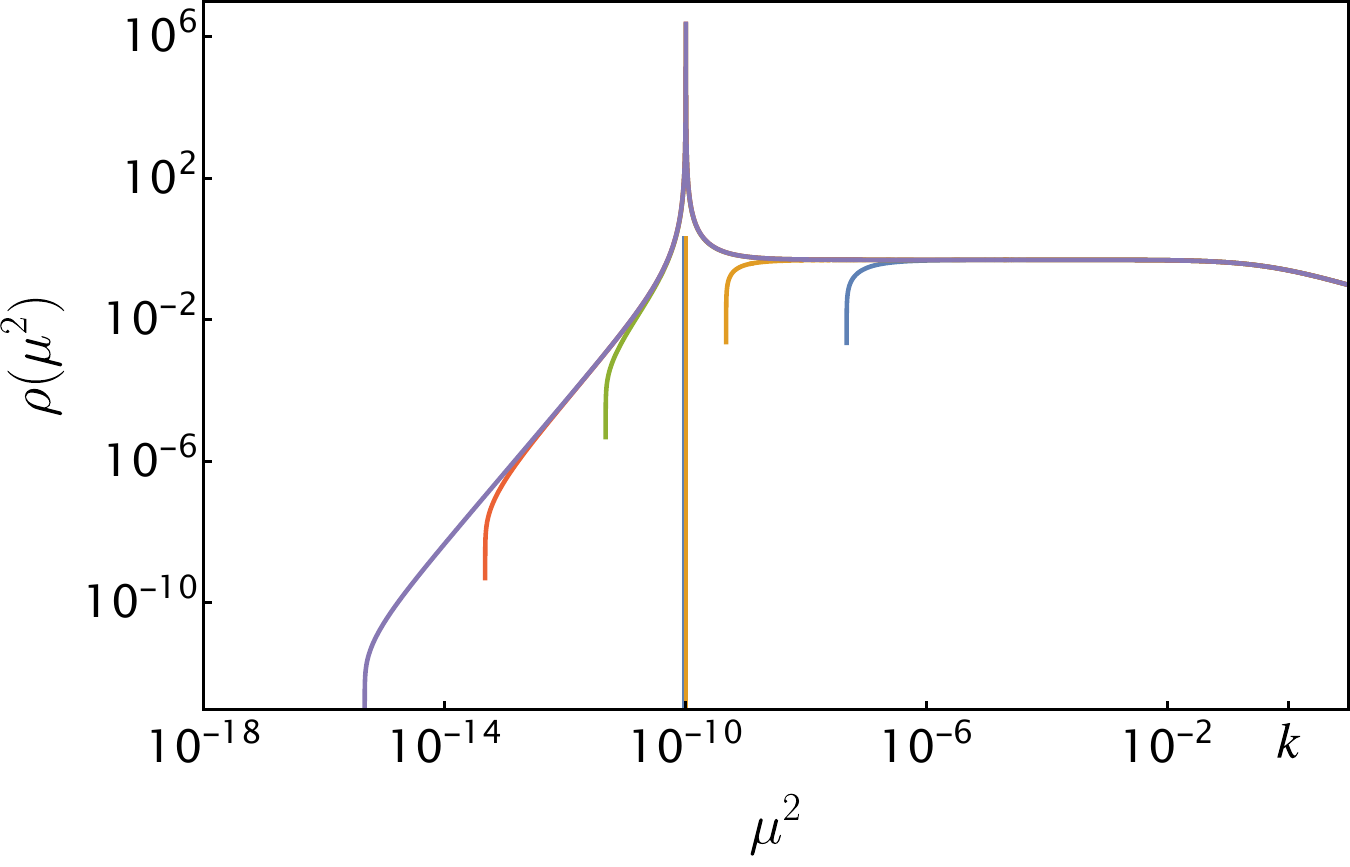}
\includegraphics[width=0.45\textwidth]{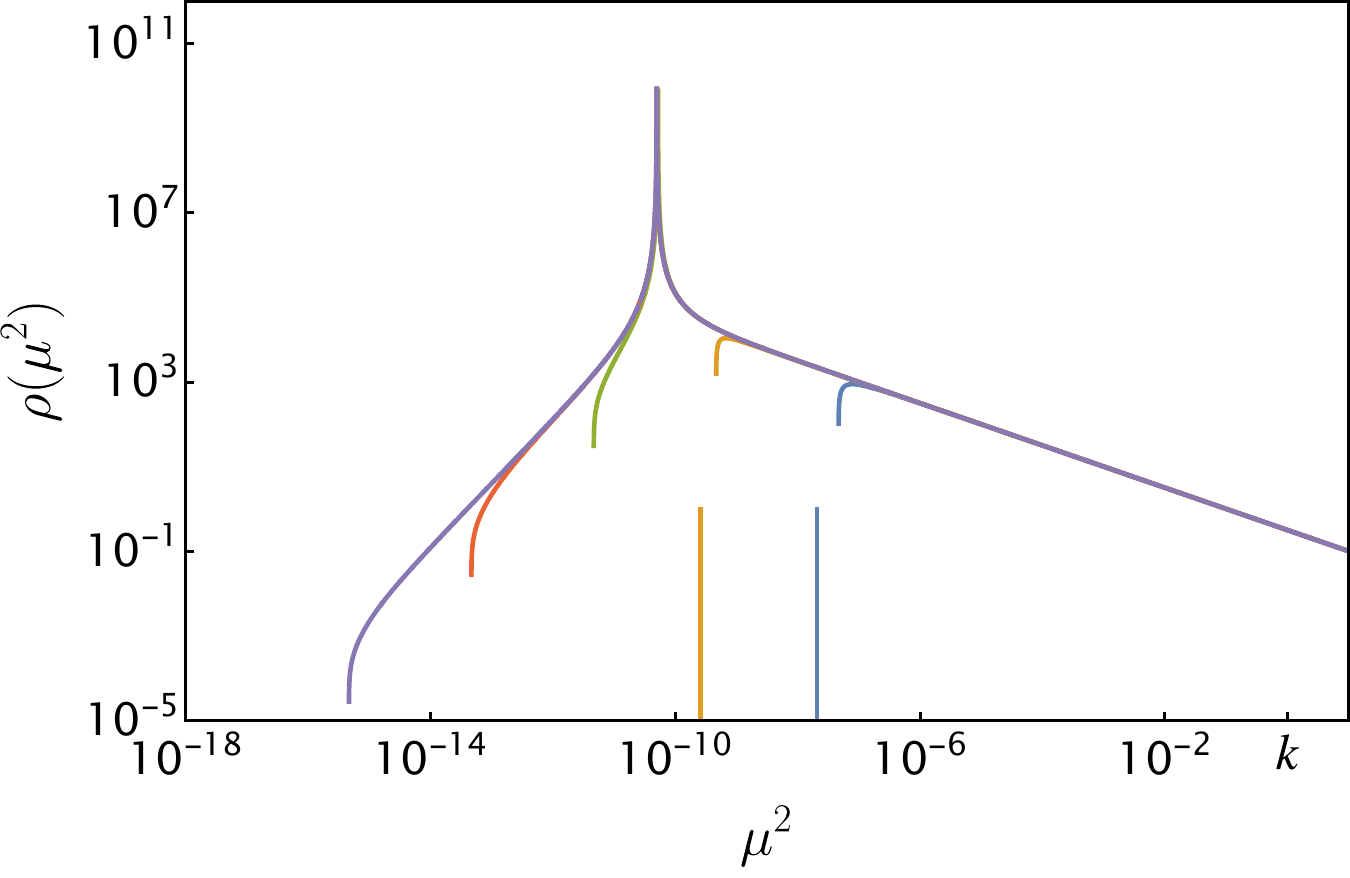}
\includegraphics[width=0.45\textwidth]{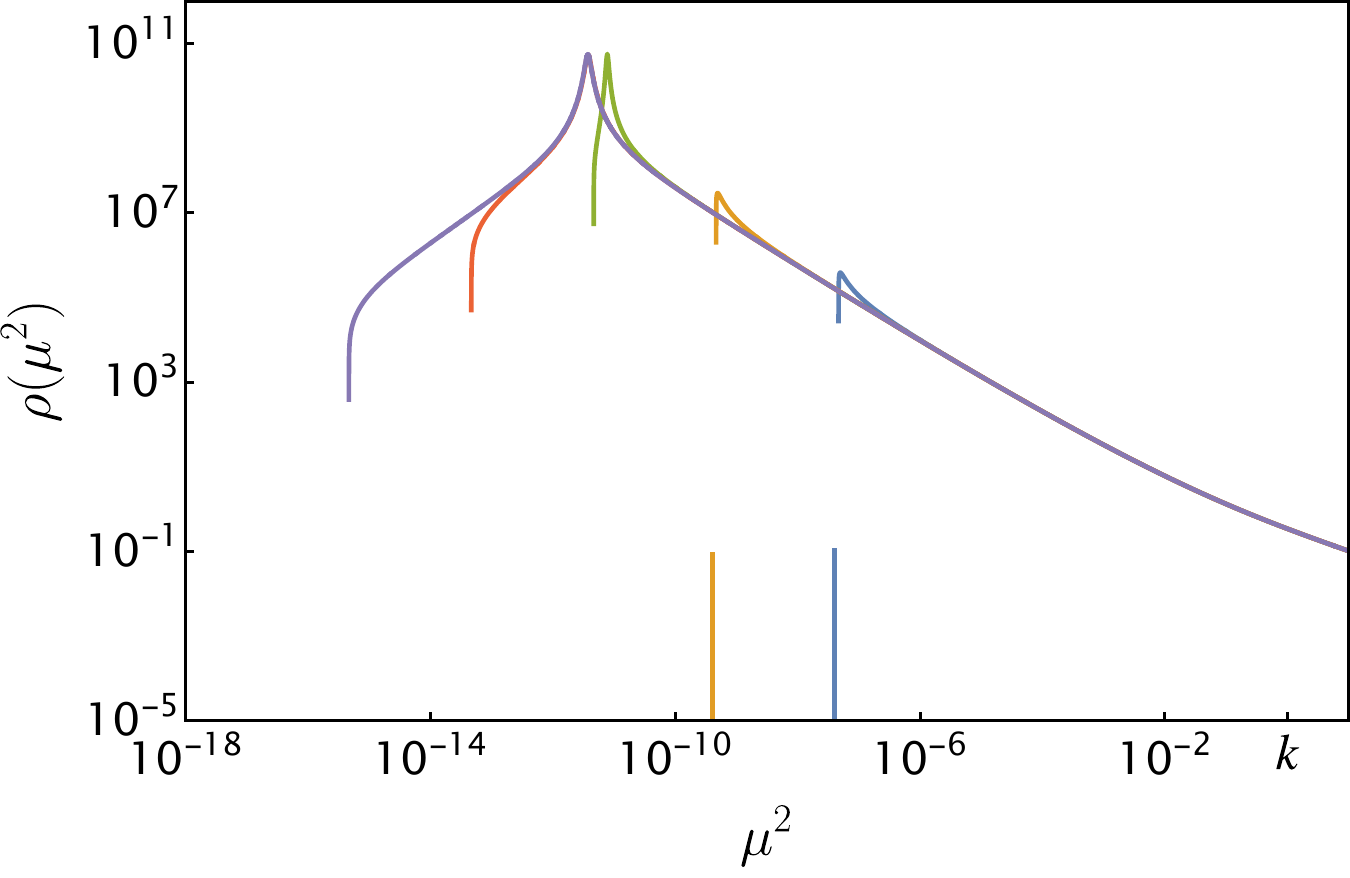}
\includegraphics[width=0.45\textwidth]{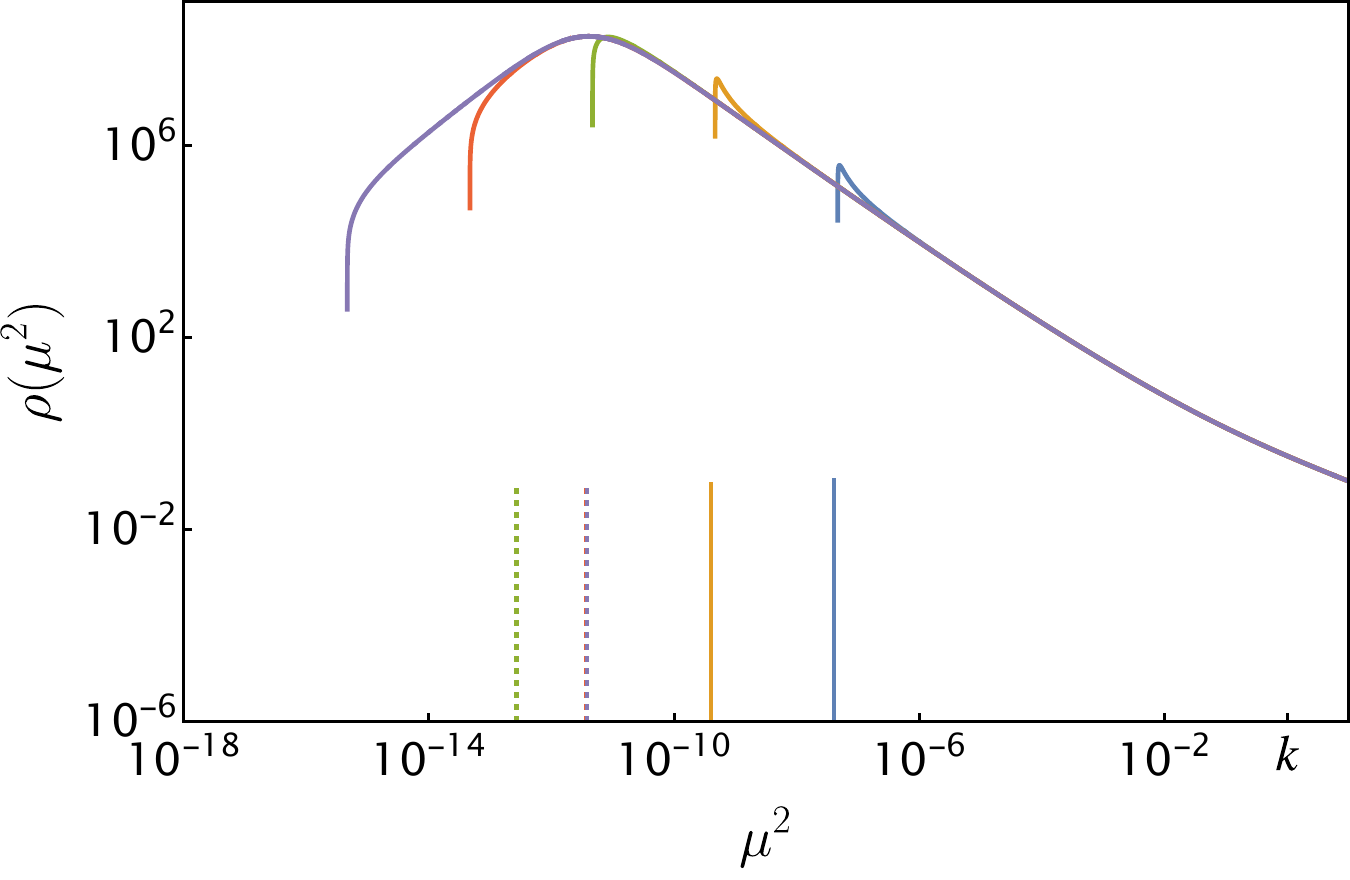}
\includegraphics[width=0.45\textwidth]{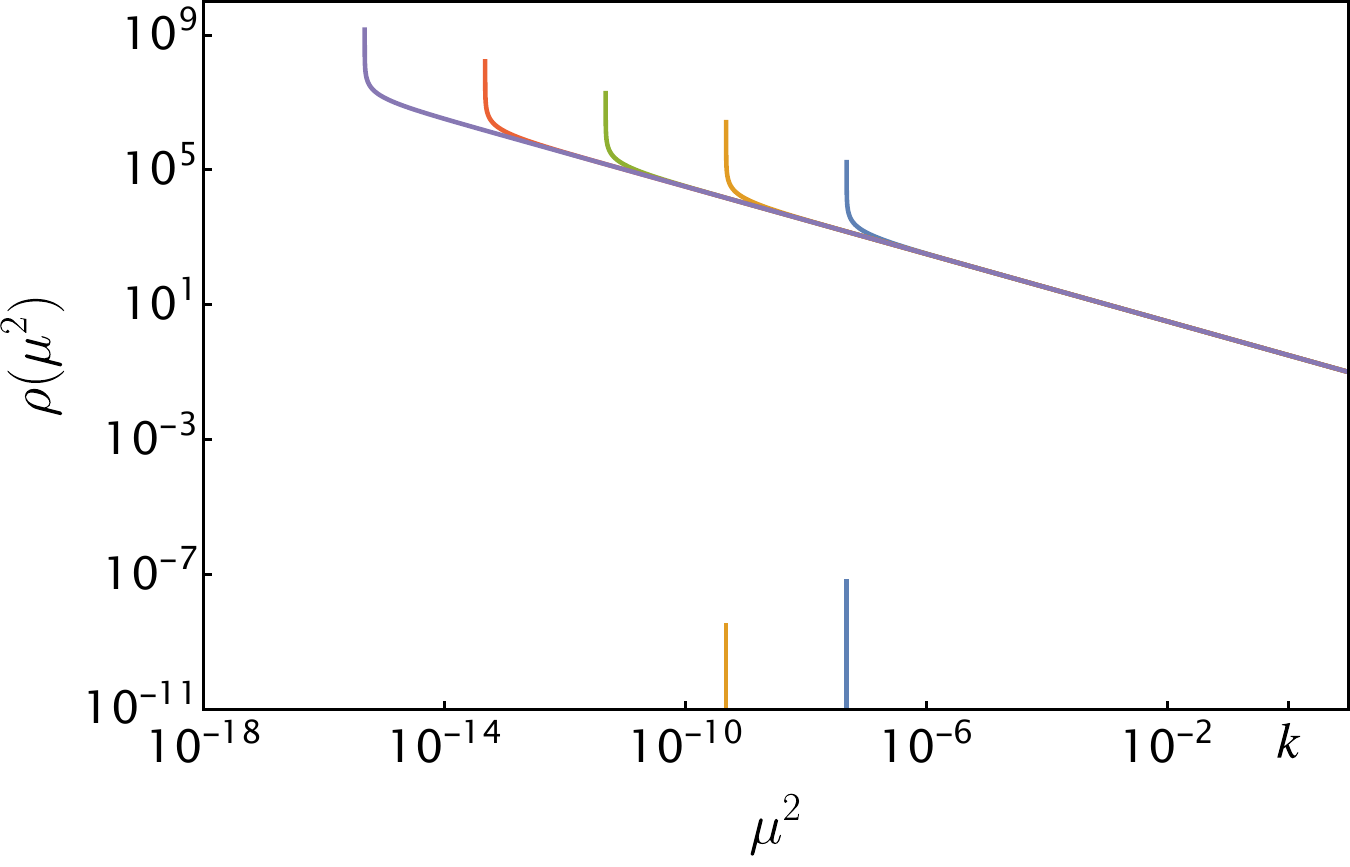}
\includegraphics[width=0.45\textwidth]{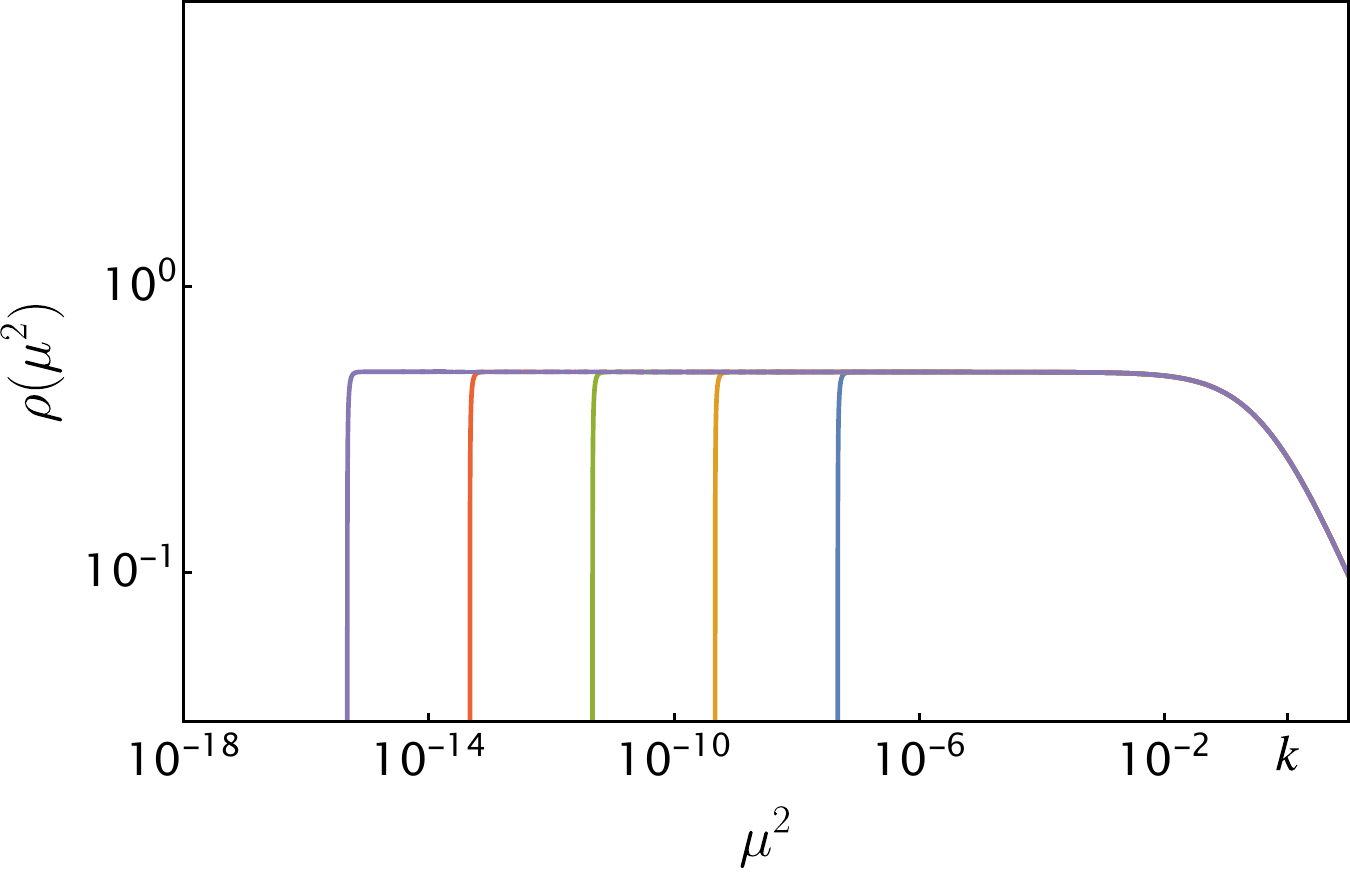}
\includegraphics[width=0.51\textwidth]{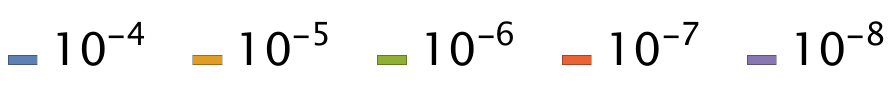}}
\caption{The spectral density for various values of the 5D parameters. In reading order, $\nu\approx 2, 1.5, 1, 1, 0.5, 0$. The boundary mass $m_0^2=2(2-\nu)+\delta$, with $\delta>0$ for all panels except plot 4. In each panel, the Hubble rate is varied according to the legend at the bottom, where $H$ is expressed in units of $k$. The vertical lines represent the light mode. The height of these lines corresponds to the coefficient, $C(\nu,H)$ of Eq.~(\ref{eq:rhoform}), in front of the delta function that appears in the spectral density below the gap.}
\label{fig:SpecDensFig}
\end{figure}

The behavior of the spectral density, $\rho(\mu^2)$, should be consistent with this interpretation.  Our expectation is therefore that $\rho$ evolves continuously from a scaling law according to the dimension $\Delta_-$ for large $\mu^2$, to the dimension $\Delta_+$ scaling in the IR, or small $\mu^2$.  If $\delta m_0^2$ is tuned to be very small, the UV scaling law $\Delta_-$ will cover a wider range of $\mu^2$ before it switches over to the $\Delta_-$ scaling dimension.  In the last 4 panels of Figure~\ref{fig:SpecDensFig}, we show the behavior of the spectral density for $\nu \lesssim 1$

We now discuss the other region, $m^2> -3$.  This situation is not as well-studied.  In this case, the standard $\Delta_-$ scaling is prohibited by unitarity constraints: $\Delta \ge 1$.  If $m^2 > -3$, the unitarity bound should somehow be enforced in the 5D theory, and the spectral density should not exhibit the forbidden $\Delta_- < 1$ scaling.  One possibility is that conformal invariance is completely broken at high scales where the theory would have followed the $\Delta_- < 1$ scaling law.  Another option is that the theory is still conformal, but exhibits a scaling law that differs from the prediction of Eq.~(\ref{eq:scalings}).  We give evidence in this section that the latter option is the correct one.

When $\nu > 1$, we find that the UV behavior of the spectral density exhibits a new scaling law $\Delta_\text{UV} = \nu$, rather than $\Delta_\text{UV} = \Delta_- = 2 - \nu$. As the theory transitions between the usual IR scaling, $\Delta_\text{IR} = \Delta_+$, and $\Delta_\text{UV} = \nu$, we find that there is typically a sharp particle-like feature in the spectral density separating the two regions of distinct scalings.~\footnote{In the complex plane, the continuum corresponds to a branch cut in the two-point function along the real axis.  The particle-like feature corresponds to a pole that is displaced above the real axis.}. 

We finally note that at very high scales, $\mu \gtrsim k$, the spectral density scales as though the extra dimension were flat, and $\rho(\mu^2) \propto (\mu^2 - m^2)^{-1/2}$~\cite{Csaki:2021gfm}.

\begin{figure}[h!]
\center{
\includegraphics[width=0.45\textwidth]{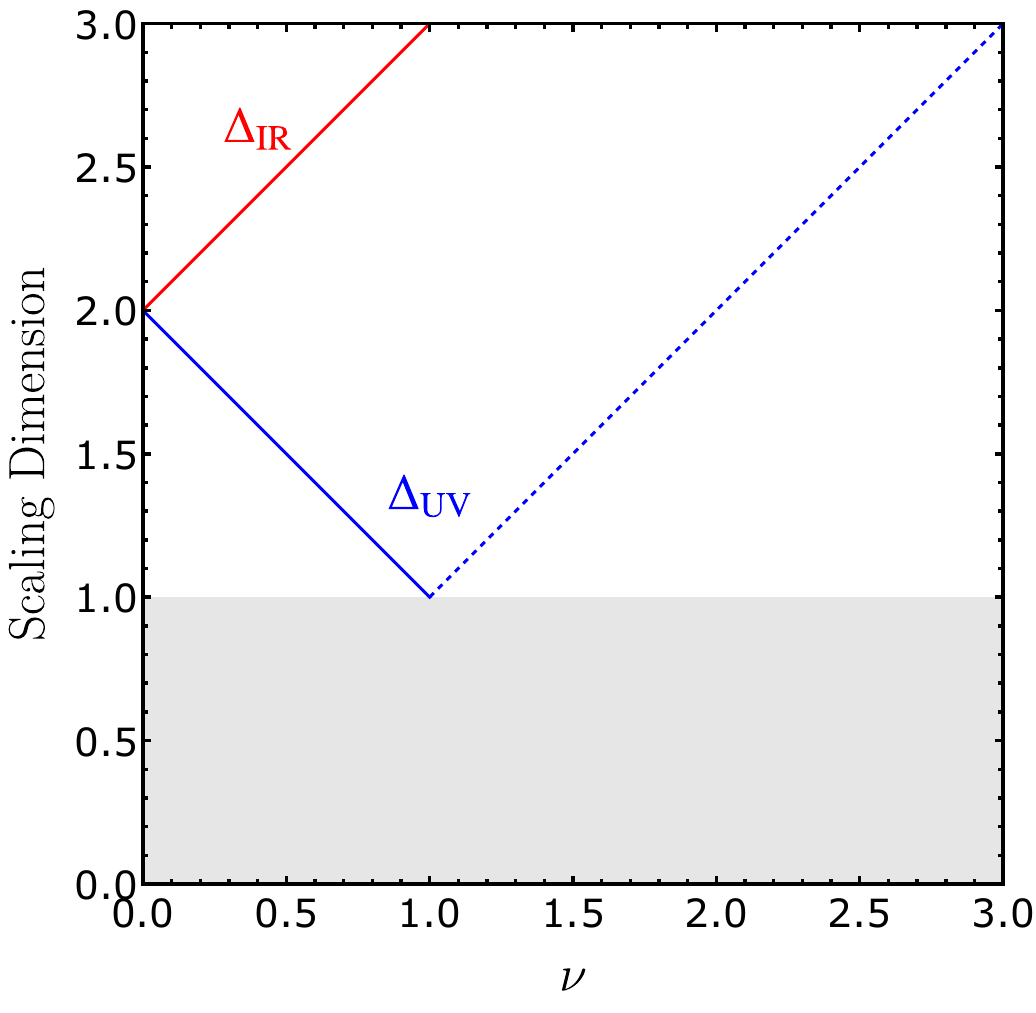}}
\caption{The scaling dimension associated with the behavior of the spectral density associated with the brane-to-brane correlator.  For $\nu < 1$, we observe the expected $\Delta_\pm$ scaling laws (solid red/blue lines), while for $\nu > 1$, we find that the $\Delta_-$ scaling law (which would have violated unitarity constraints) is replaced with a novel holographic scaling law $\Delta_\text{UV} = 2-\Delta_- = \nu$ (the blue dashed line).}
\label{fig:ScalingDims}
\end{figure}

\begin{figure}[h!]
\center{
\includegraphics[width=0.49\textwidth]{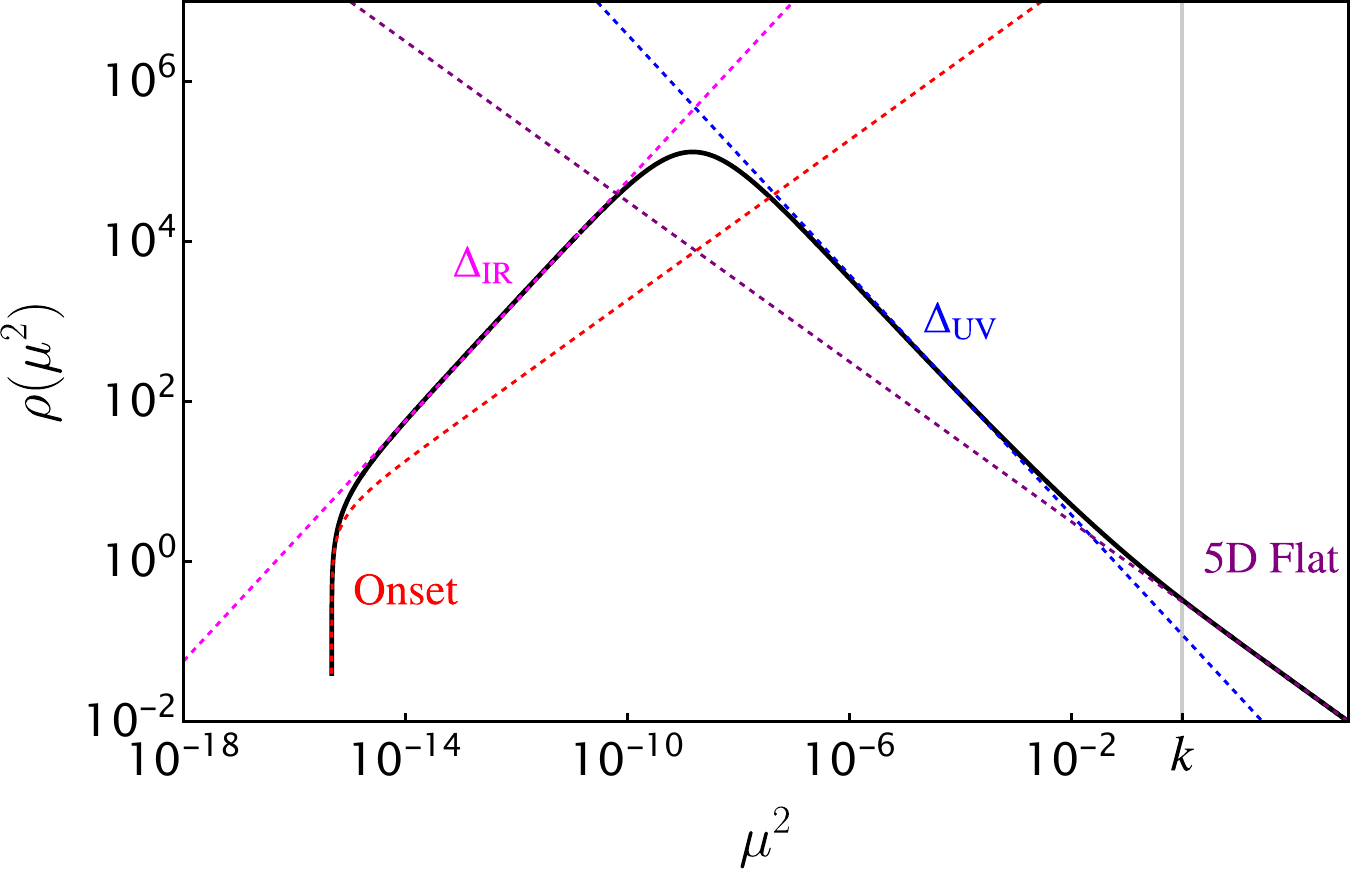}
\includegraphics[width=0.49\textwidth]{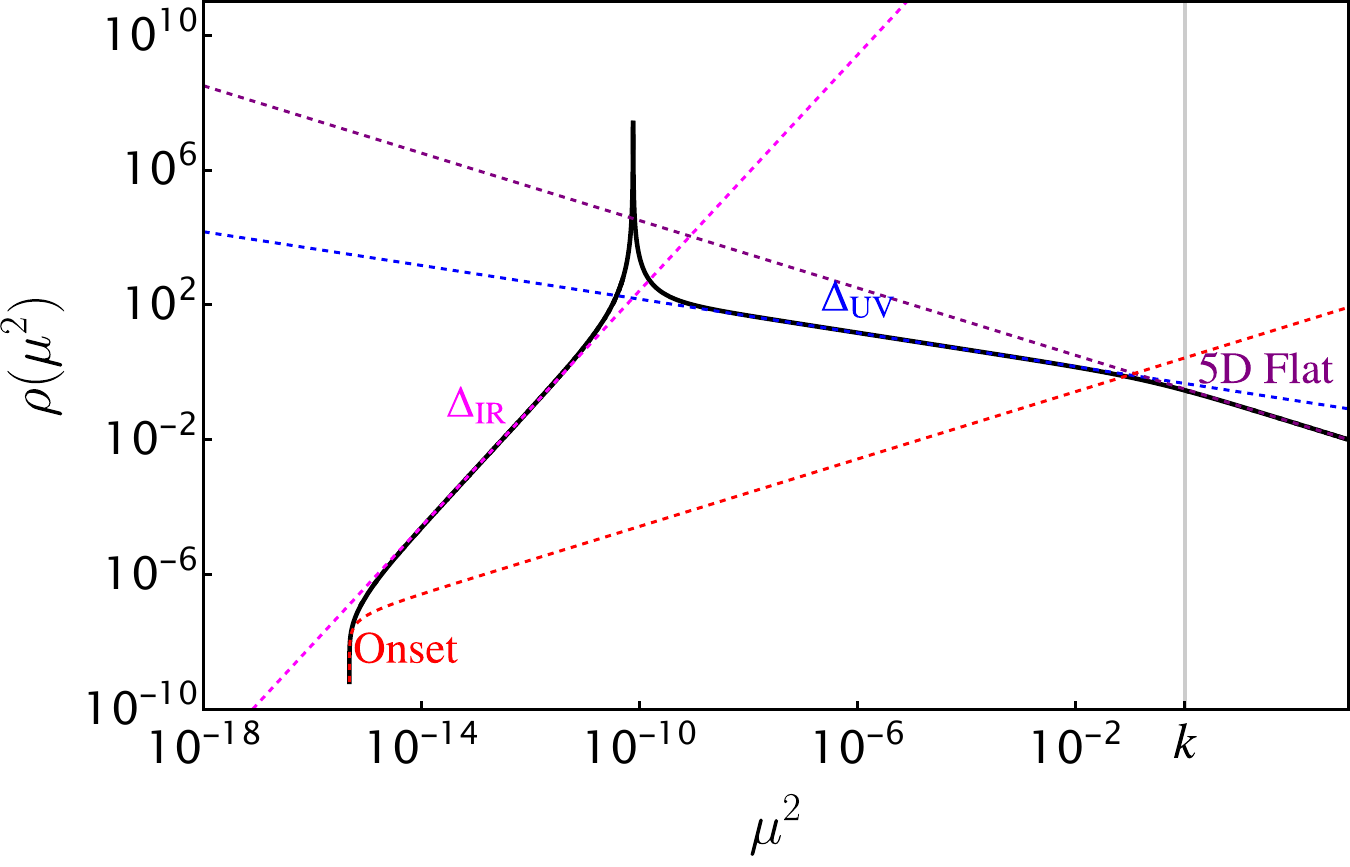}}
\caption{In these plots, we display the different regimes of scaling behavior of the spectral density.  These regimes are divided into the onset, the IR scaling region, the UV scaling region, and the 5D scaling region.  The onset region is characterized by a spectral shape given by $\sqrt{\frac{\mu^2}{H^2} - \frac{9}{4}}$.  The IR behavior is determined by a scaling dimension $\Delta_\text{IR} = \Delta_+$.  The UV behavior is given by a scaling with dimension $\Delta_\text{UV} = 2 - \nu$ if $\nu <1$ (left), and by $\Delta_\text{UV} = \nu$ if $\nu > 1$ (right).}
\label{fig:SpecDensAnatomy}
\end{figure}

In Figure~\ref{fig:ScalingDims}, we display a plot of the observed scaling behavior of the two point function encoded in the spectral density, and in Figure~\ref{fig:SpecDensAnatomy} we display the basic features of the spectral density in the two situations $\nu < 1$ and $\nu > 1$.

As a function of the parameter $\nu$, then, we can consider the theory to be exhibiting a phase transition as $\nu$ passes through $1$.  Indeed, in the cosmological $H \ne 0$ context,  the spectral density can be characterized, for $\nu >1$, as consisting of a delta function contribution (the single particle state), with a mass of $\mu_*^2 = (\nu -1) \left[ m_0^2 - 2 (2-\nu)\right] + 2 (2-\nu) H^2$, and a continuum beginning at $\bar{\mu}^2 = \frac{9}{4} H^2$.  As $\nu$ passes through $1$ from above, then, the particle mass approaches $\mu_*^2 = 2 H^2$, and must undergo some sort of fate when the $\Delta_-$ scaling again becomes viable, and the particle solution is no longer normalizable.

The spectral density can therefore roughly be characterized as
\beq
\rho(\mu^2) = C(\nu,H) \delta (\mu^2-\mu_*^2) + \rho_c (\nu, m_0, \mu^2,H) \Theta \left(\mu^2 - \frac{9}{4} H^2 \right).
\label{eq:rhoform}
\eeq
In this spectral density, we have a potential delta-function contribution from a particle with mass $\mu^*$, as well as the continuum beginning at $\bar{\mu}$.  The spectral density in 5D is calculated by considering the brane-to-brane propagator, which provides explicit functions for $C$ (the residue associated with the simple pole at $p^2 = \mu_*^2$) and for $\rho_c$ (the spectral density of the continuum alone).  

%%%%%%%%%%%%%%%%%%%%%%%%%%%%%%%%%%%%%%%%%%%%%%%%%%%%
%%%%%%%%%%%%%%%%%%%%%%%%%%%%%%%%%%%%%%%%%%%%%%%%%%%%
%%%%%%%%%%%%%%%%%%%%%%%%%%%%%%%%%%%%%%%%%%%%%%%%%%%%
%%%%%%%%%%%%%%%%%%%%%%%%%%%%%%%%%%%%%%%%%%%%%%%%%%%%

\section{Cosmological Collider Observables}
\label{sec:coscoll}

In our simplified model, we are examining the case of a UV brane localized scalar inflaton.  Expanding about the background of its slow-rolling vacuum expectation value, we have $\phi(t,x) = \phi_0(t) + \xi(t,x)$.  Utilizing the gauge freedom of 4D gravity, the curvature fluctuation is in turn can be expressed as $\zeta = -\frac{H}{\dot{\phi}_0} \xi$~\cite{Maldacena:2002vr}.  Note that this is in the context of the low-energy 4D theory, and is valid, in our case, at momenta below the Hubble rate.  Otherwise, the curvature fluctuation must be diagonalized and its 5D wavefunction fully identified.  The correlators for the field $\zeta$ are the gauge invariant observables.

Mixing of the 5D scalar field with the curvature perturbation is suppressed, as the degrees of freedom in associated with it are gapped at scales near $H$.  The approximation of single field inflation used above is thus a good one, and we can treat corrections to it (such as the 3-point function we are most interested in for this work) perturbatively.

The scalar CFT operator (holographically speaking, the bulk scalar field) gives a contribution to the 4-point function for $\xi$, and thus a non-trivial 3-point correlator for $\zeta$, when one of the legs is taken to be the background value for the inflaton.  We presume that the interaction between the inflaton and the bulk scalar field on the UV brane is a derivative interaction, respecting an approximate shift symmetry of the the inflaton field:
\beq
{\mathcal S}_\text{int} = \lambda \int_{z=z_0} d^4 x \sqrt{g} \Phi (\partial \phi)^2.
\eeq
This is a dimension-11/2 operator, since the bulk field $\Phi$ is dimension 3/2.  Thus $\lambda$ carries mass dimension $-3/2$.  In this paper, we work in units of the curvature parameter, $k$.  Typically, one would expect $\lambda M_\text{Pl}^{3/2}$ to be ${\mathcal O}(1)$.  Since $k$ is taken to be somewhat small to generate a sizeable hiearchy, and to have a perturbative 5D gravity theory, $\lambda$ is then expected to be $\lambda \sim {\mathcal O}(30-100)$ in units of $k$, utilizing naive dimensional analysis~\cite{Georgi:1992dw}.  

The 3-point correlator corresponding to the exchange of a single massive particle with mass $m^2 = H^2 (\gamma^2 +9/4)$, with external momenta $\vec{k}_{1,2,3}$ in the ``squeezed'' limit $k_3 \ll k_{1,2}$, is then given by~\cite{Arkani-Hamed:2015bza}
\beq
\langle \xi_{\vec{k}_1} \xi_{\vec{k}_2} \xi_{\vec{k}_3} \rangle_\gamma = \dot{\phi}_0(\tau^*) \frac{\lambda^2}{2 k_1^2 k_3^2} J(\gamma, k_3/k_1),
\eeq
where 
\begin{align}
J(\gamma, k_3/k_1) =& \frac{1}{4 \sqrt{\pi}} \frac{\pi^2}{\cosh^2 \pi \gamma} \left( \frac{k_3}{k_1} \right)^{3/2} \times \nonumber \\
& \left[ \left( \frac{k_3}{4k_1} \right)^{-i \gamma} \frac{(1-i \sinh \pi \gamma) (5/2-i \gamma) (3/2 - i \gamma) \Gamma(i \gamma)}{\Gamma(1/2+i \gamma)} + \text{h.c.} \right]
\label{eq:Jfunc}
\end{align}
For modes with mass above the gap, a ``smoking gun" signal of oscillations with wavenumber $\gamma$ as a function of $\log k_3/k_1$, damped by a universal $(k_3/k_1)^{3/2}$ scaling, parametrize the spectrum.  For modes with mass below the gap, $\gamma$ is imaginary, and the leading behavior is a characteristic scaling behavior $(k_3/k_1)^\alpha$, with $\alpha = 3/2 - | \gamma |$.

This is for a single massive particle.  In the scenario we are studying, there is a continuum of modes above the gap, and there may or may not be (depending on the UV boundary conditions and bulk mass) a particle contribution as well.

To obtain the corresponding result for the 5D theory, we must convolute the single particle contribution with the spectral density associated with the UV-UV brane propagator.  In the squeezed limit, this is given by:
\beq
\langle \xi_{\vec{k}_1} \xi_{\vec{k}_2} \xi_{\vec{k}_3} \rangle' = \dot{\phi}_0(\tau^*) \frac{\lambda^2}{2 k_1^2 k_3^2} \int_0^\infty d\mu^2 \rho(\mu^2) J(\gamma, k_3/k_1).
\eeq
Note that we have dropped the usual momentum conserving delta-function.
Finally, to compare with experiment, we normalize this result by the two-point function, and express it in terms of the physical curvature perturbation, $\zeta$:
\beq
\frac{\langle \zeta_{\vec{k}_1} \zeta_{\vec{k}_2} \zeta_{\vec{k}_3} \rangle'}{4\langle \zeta_{\vec{k}_1} \zeta_{-\vec{k}_1}\rangle' \langle \zeta_{\vec{k}_3} \zeta_{-\vec{k}_3}\rangle'} = - \epsilon M_\text{Pl}^2 \lambda^2 \int_0^\infty d\mu^2 \rho(\mu^2) J(\gamma, k_3/k_1)=  \frac{3}{5} F_\text{NL} (k_3/k_1) .
\eeq
Where in the last equality, we have expressed the correlator in terms of the standard parametrization, $F_\text{NL}$.  Commonly, constraints on $F_\text{NL}$ are given for a magnitude multiplying a shape function, $S$:
\beq
F_\text{NL} (k_3/k_1) = f_\text{NL} \cdot S(k_3/k_1),
\eeq
where the shape function, $S$ is typically normalized to $S(1) = 1$.  For our purposes, we utilize a slightly different normalization, since we are working in the squeezed limit where $k_3/k_1 \ll 1$.  Also, in our study, the scalar spectral densities are all gapped at scale $\bar{\mu} = \frac{3}{2}H$, and from Eq.~\ref{eq:Jfunc}, we see that the continuum of modes above $\bar{\mu}$ will contribute with an overall power law of $(k_3/k_1)^{3/2}$.  For this reason, it is useful to factor out that scaling that will be universal to most regions of parameter space.  We thus write
\beq
F_\text{NL} (k_3/k_1) = f_\text{NL} \left( \frac{k_3}{k_1} \right)^{3/2} \tilde{S}(k_3/k_1),
\label{eq:FNLeq}
\eeq
and we report results for the bispectrum calculations in terms of $f_\text{NL}$, and a corresponding \emph{modified} shape function, $\tilde{S}$. We choose our normalization so that, over the range $10^{-6} < k_3/k_1 < 10^{-1}$, the maximum magnitude of $\tilde{S}$ is 1:  $|\tilde{S}|_\text{max} = 1$.  This corresponds to a value of $k_3/k_1$ that is in the squeezed regime, and also ensures that an oscillatory signal, if present, has magnitude near 1.  This prevents phase information in the signal from becoming convoluted into the definition of $f_\text{NL}$.

\begin{figure}[h!]
\center{
\includegraphics[width=0.45\textwidth]{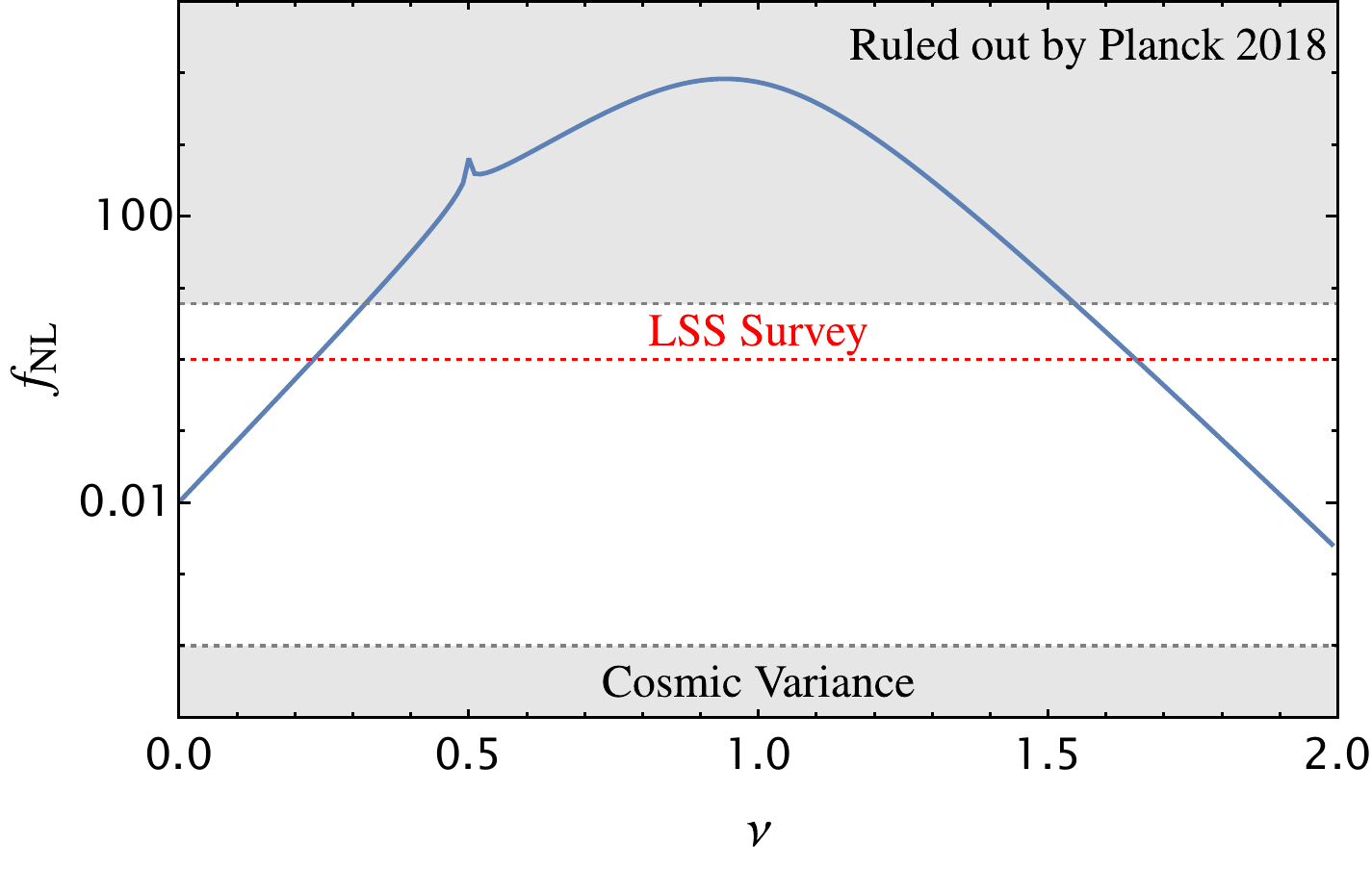}
\includegraphics[width=0.44\textwidth]{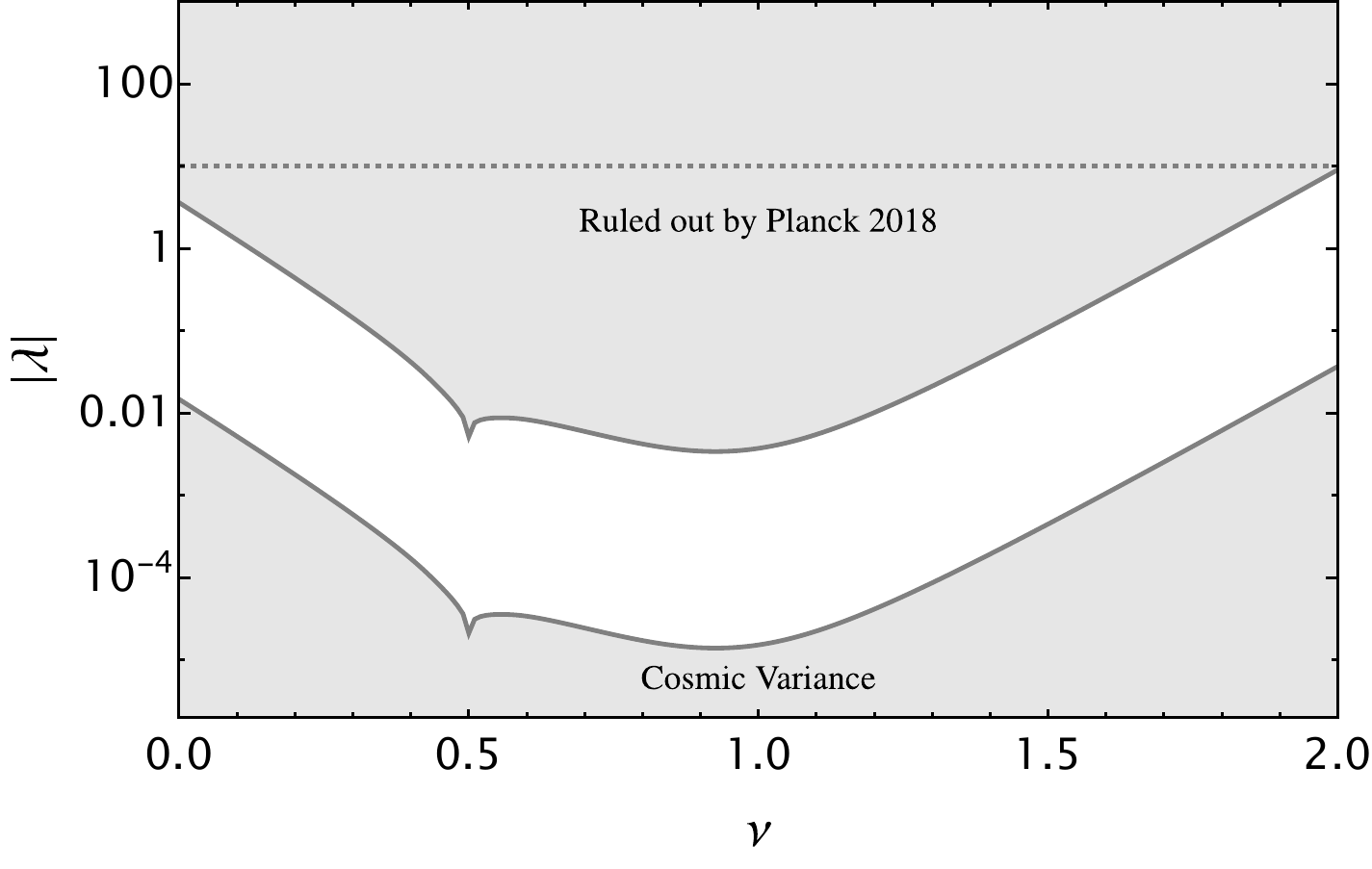}
}
\caption{In the figure on the left, we report on the value of $f_\text{NL}$ as defined in Eq.~(\ref{eq:FNLeq}) for various values of $\nu$.  We have presumed that the inflationary Hubble rate is at its maximum allowed value.  We have taken $\delta = 0$, and we have also taken $\lambda = 1$.  The shaded region above is roughly excluded by Planck 2018 data~\cite{Planck:2018jri}.  The shaded region below is a floor that cannot be probed due to the limitations imposed by cosmic variance.  In the figure on the right, we plot the same data, but we interpret the ceiling and floor as bounds on $\lambda$.}
\label{fig:CCPFig}
\end{figure}

\begin{figure}[h!]
\center{
\includegraphics[width=0.45\textwidth]{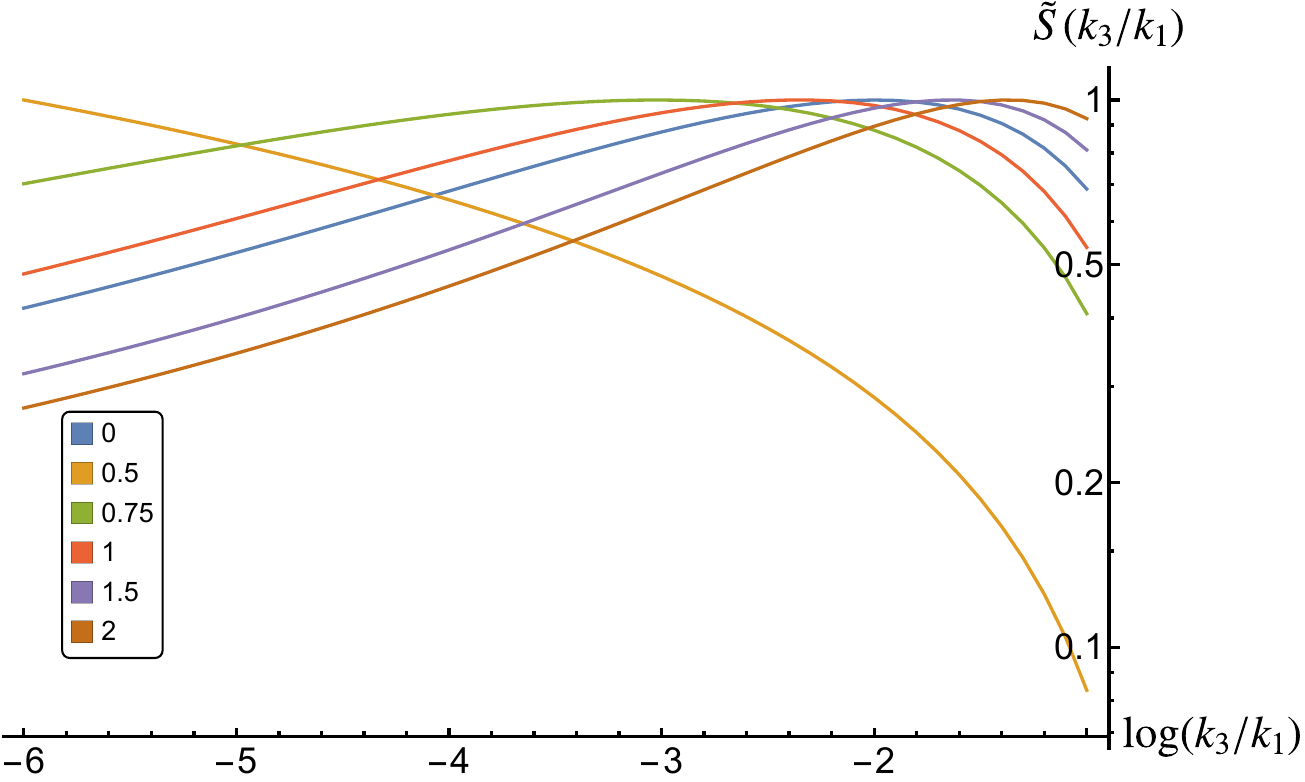}
\includegraphics[width=0.44\textwidth]{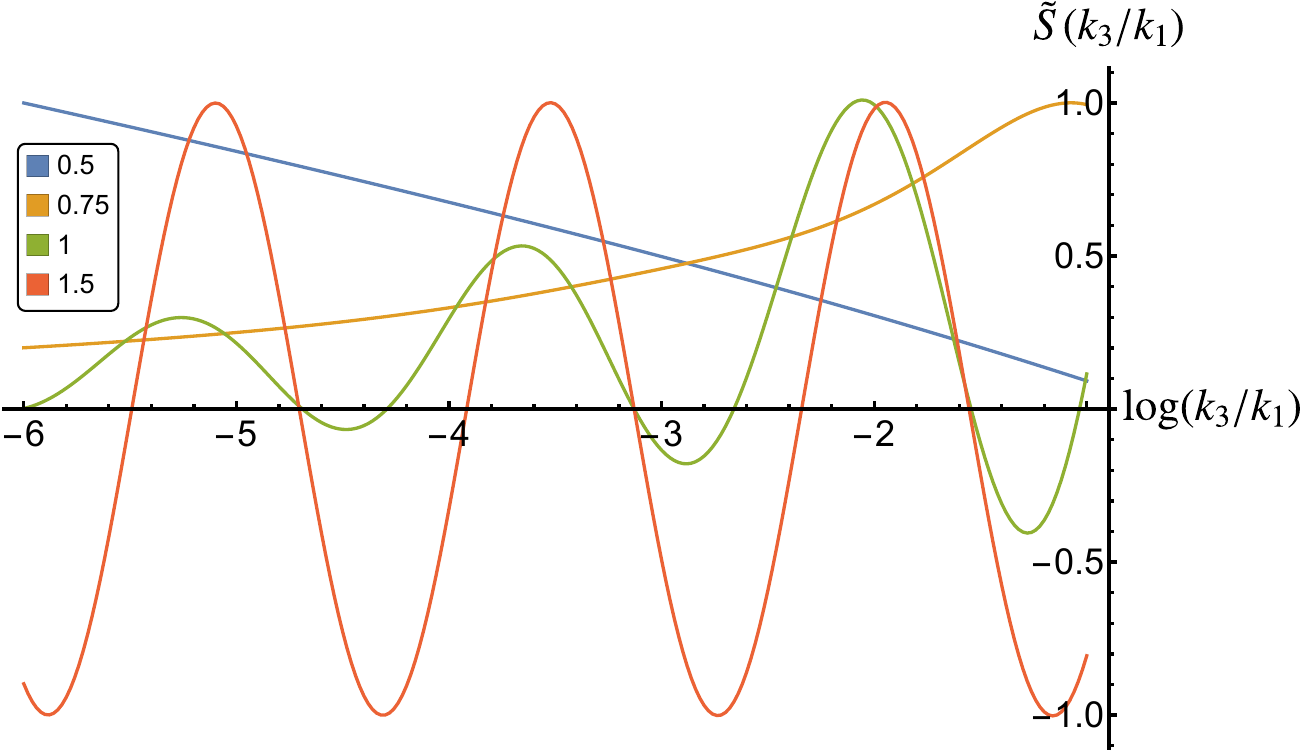}
}
\caption{In these plots, we display the modified shape function, $\tilde{S} = S \cdot (k_3/k_1)^{3/2}$.  With the dominant scaling removed, it is easier to distinguish the shapes corresponding to different values of the 5D parameters.  In the plot on the left, we have taken $\delta = 0$, corresponding to a fine-tuned (or supersymmetric) boundary condition.  In this case, the $\Delta_\text{UV}$ scaling sets in soon after the onset of the continuum.  In the plot on the right, we have tuned parameters so that there is a particle-like feature in the spectral density just above the onset of the continuum, at $\gamma = \sqrt{3}$.  We note the presence of ``smoking gun" oscillatory features on top of a continuum contribution, which can cause the oscillations to be displaced.}
\label{fig:ShapeFig}
\end{figure}

In Figures~\ref{fig:CCPFig} and~\ref{fig:ShapeFig}, we display calculations of $f_{NL}$ for various values of the parameters.  We note that there is a considerable increase in the value of $f_\text{NL}$ at $\nu \approx 1/2$.  This coincides with a transition where a particle-like feature moves from the continuum into the region below the gap.  When there is a sharp particle-like feature close to (but slightly above) the gap, the modes corresponding to that particle can cause a particular frequency in this superposition to dominate, leading to a striking oscillatory feature in the bispectrum.  In these figures, we also indicate a rough idea of existing constraints from Planck data.  These are typically overly conservative, in that we only compare the squeezed limit calculation of $f_\text{NL}$, without considering contributions to the equilateral shape non-gaussianities that typically accompany the squeezed features (See, e.g.~\cite{Qin:2023ejc} for general features of non-gaussianites, or~\cite{Kumar:2018jxz}, for a comparison in a similar model).

We draw attention in particular to features of these models that distinguish their phenomenology from that of heavy particles near the threshold, $3/2 H$.  First, all shape functions that we observe exhibit the characteristic universal $(k_3/k_1)^{3/2}$ scaling, typically without oscillations.  This arises from superposing the numerous modes in the continuum that all contribute oscillations, though at different frequency.  Second, we note that when parameters are chosen such that there is a particle slightly above this threshold, there are signals that are unique to that of a continuum with a particle-like feature in the spectral density.  For example, the green line in the plot on the right side of Figure~\ref{fig:ShapeFig} displays oscillations that are displaced from the origin, corresponding to the convolution of a continuum contribution along with a particle contribution.

%%%%%%%%%%%%%%%%%%%%%%%%%%%%%%%%%%%%%%%%%%%%%%%%%%%%
%%%%%%%%%%%%%%%%%%%%%%%%%%%%%%%%%%%%%%%%%%%%%%%%%%%%
%%%%%%%%%%%%%%%%%%%%%%%%%%%%%%%%%%%%%%%%%%%%%%%%%%%%

\section{Inflation and Radius Stabilization}
\label{sec:rad_stab}
Stabilization of the Planck-weak hierarchy in Randall-Sundrum models is required to obtain acceptable phenomenology, and it is interesting to consider whether there can be cosmological signatures of a Goldberger-Wise stabilizing scalar field that are unique to the regimes of parameter space relevant for these models~\cite{Goldberger:1999un}.  

The stabilization mechanism corresponds typically to a 5D scalar field which gets a vacuum expectation value, which in turn deforms the geometry due to gravitational backreaction effects.  This deformation of the geometry leads to the generation of a potential for the radion, a particle degree of freedom corresponding to perturbations of a 2nd brane, the IR brane, and dual to a pseudo-dilaton associated with the spontaneous breaking of approximate conformal symmetry.

The scalar action in the stabilized geometry with an IR brane is given by
\beq
S = \int d^5 x \sqrt{g} \left[ (\partial_M \Phi )^2 - m^2 \Phi^2 \right] - \int d^4 x \sqrt{g_0} V_0 (\Phi) - \int d^4 x \sqrt{g_1} V_1 (\Phi)
\eeq
where $V_{0,1}$ are potentials on the UV and IR brane, and $g_{0,1}$ are the induced metrics on these branes.  The background geometry in this case is not AdS-dS, but rather AdS (up to backreaction effects, which we presume to be small).  The inflaton in this case is no longer rolling, but rather sitting at its potential minimum.  The effective cosmological constant is close to zero, matching current observational constraints.

In the language of the dual CFT, the GW stabilization mechanism corresponds to a small explicit breaking of conformal invariance due to the sourcing of a nearly marginal operator.  The breaking of conformal symmetry is associated with confinement, and the confinement scale arises through dimensional transmutation associated with the slow running of that nearly marginal operator.  In 5D, this corresponds to two possibilities that have been explored in the literature.  

The most commonly employed mechanism involves turning on a single-trace operator with scaling dimension close to 4, which in 5D corresponds to a 5D scalar field with a bulk mass close to zero.  A second option is to source a double-trace operator which is nearly marginal.  This corresponds to a 5D scalar with mass term close to the BF bound~\cite{Vecchi:2010em,Geller:2013cfa}, with a moderately tuned value of a UV-brane localized mass.  This model is particularly interesting since, in this model, the stabilizing scalar can also play the role of the Higgs field.  We will consider both of these scenarios in turn.

In the case of the inflating geometries we study, there is only the UV brane -- the IR brane is obscured by the horizon.~\footnote{See~\cite{Kumar:2018jxz} for a similar study of the cosmology of RS models, but where the confinement scale is close to the Hubble scale, $H$.}  This can occur when an inflaton field on the UV brane is displaced from its minimum, and is undergoing slow-roll.  This can create a Hubble rate with an approximate deSitter temperature which exceeds the confinement scale.  During inflation, then, the CFT is in the unbroken de-confined phase, though gapped due to explicit breaking of conformal invariance.  The stabilizing scalar field is still present, however, and can couple to the inflaton and give rise to non-gaussian features in the density perturbations.

Expanding in fluctuations of the GW field to quadratic order leads to an effective action with the same bulk mass as the background vev, and with an effective boundary mass.  Therefore, our analysis of the previous sections goes through effectively unchanged.  We focus now on results specific to these two classes of stabilization model.

\begin{center}
{\bf Small Bulk Mass:  $m^2 \approx 0$}
\end{center}
In this scenario, the model is dual to sourcing an operator ${\mathcal O}$ with scaling dimension dimension close to 4:
\beq
{\mathcal L} \ni \lambda {\mathcal O}.
\eeq
The coupling $\lambda$ is nearly marginal, and thus runs slowly.   Confinement thus occurs through a form of dimensional transmutation, stabilizing the Planck-Weak hierarchy, however the size of the coupling, $\lambda$ need not be especially small, unlike the QCD coupling constant at high energies.  

The 5D theory, however, should remain perturbative.  On the 5D side, the bare coupling $\lambda$ is associated with the value of the vev of the bulk scalar field on the UV boundary, which is in turn associated with the boundary conditions associated with the potential $V_0$.  The equation of motion which determines the vev profile can then be associated with the renormalization group equation for $\lambda$.

A common form chosen for the UV boundary potential is
\begin{equation}
V_0 = \Lambda_0 + \frac{1}{2} M^2 \left( \phi - \phi_0 \right)^2.
\end{equation}
In the existing literature, a simplifying assumption is commonly taken: $M^2 \rightarrow \infty$ (or the ``stiff-wall" limit).  Taking this limit sets the value of the scalar field on the UV brane to be $\phi(y=0) = \phi_0$.  We avoid this limit since it gives the fluctuations an infinite boundary localized mass term, and forces them to vanish on the UV boundary.  This would in turn set the interaction strength between this field and the inflaton to zero.  For finite $M^2$, we identify $M^2 = m_0^2$, in the notation of the previous sections.  We thus focus on the phenomenology of $\nu$ close to 2, while varying over values of $\delta$.

\begin{figure}[h!]
\center{
\includegraphics[width=0.48\textwidth]{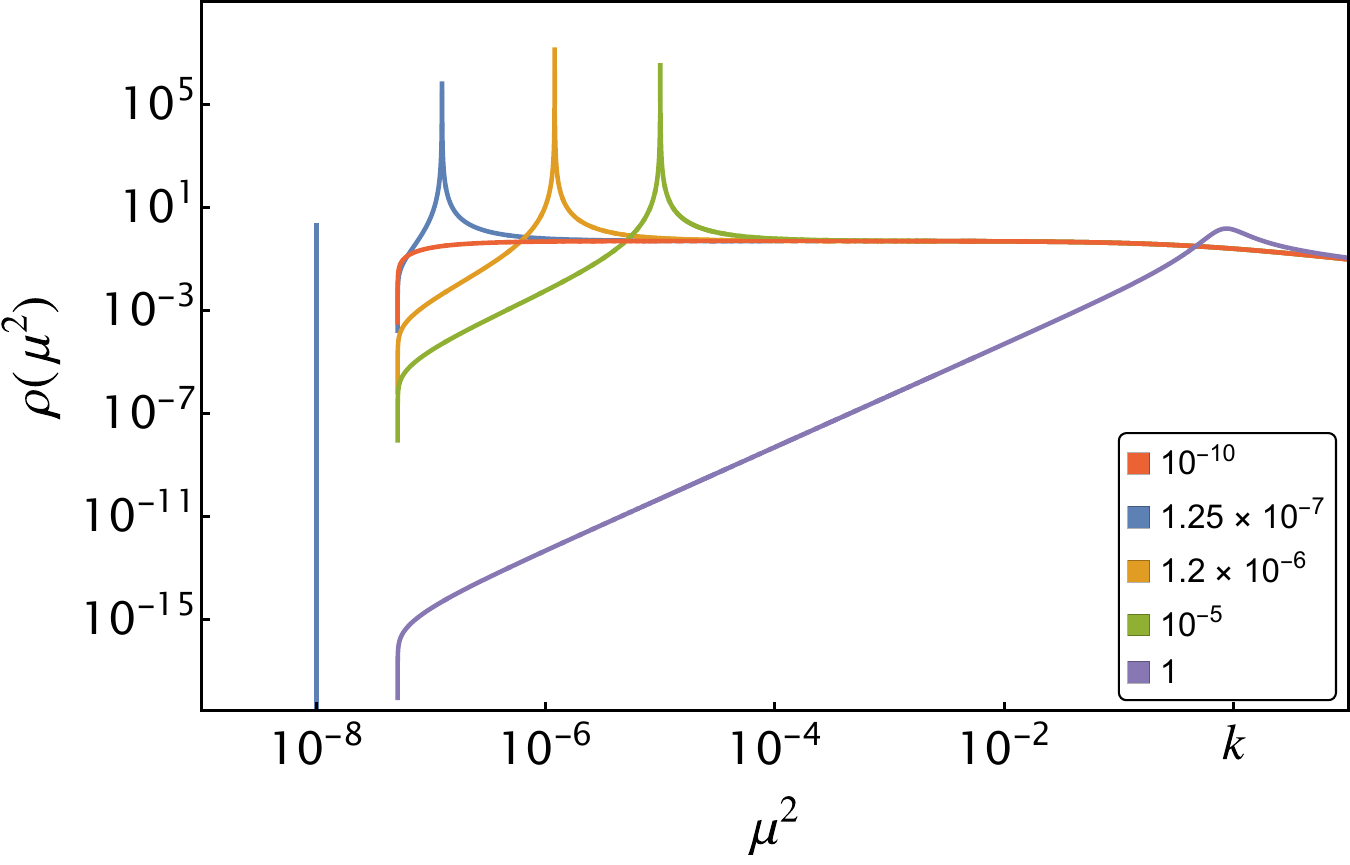}
\includegraphics[width=0.48\textwidth]{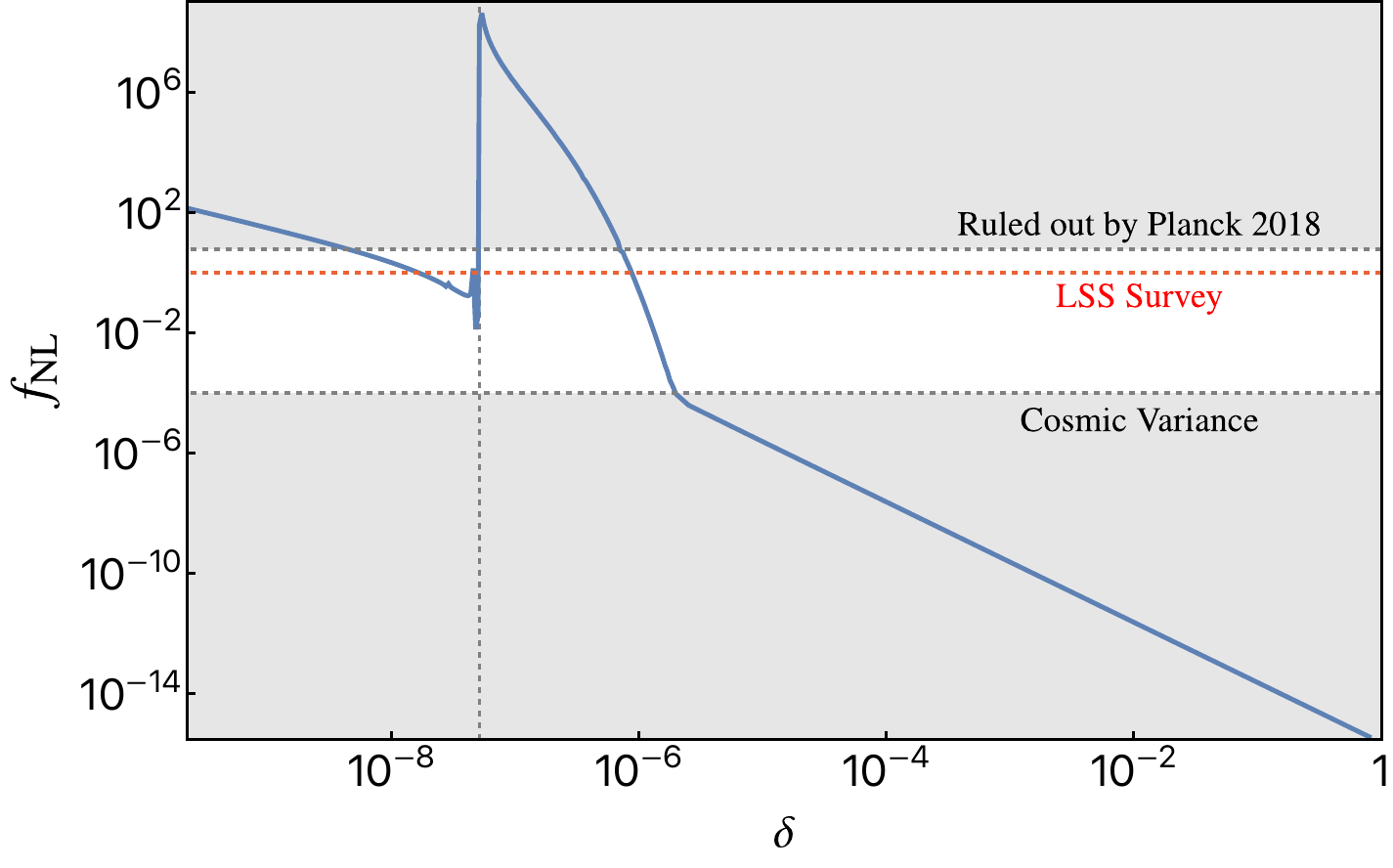}
\includegraphics[width=0.48\textwidth]{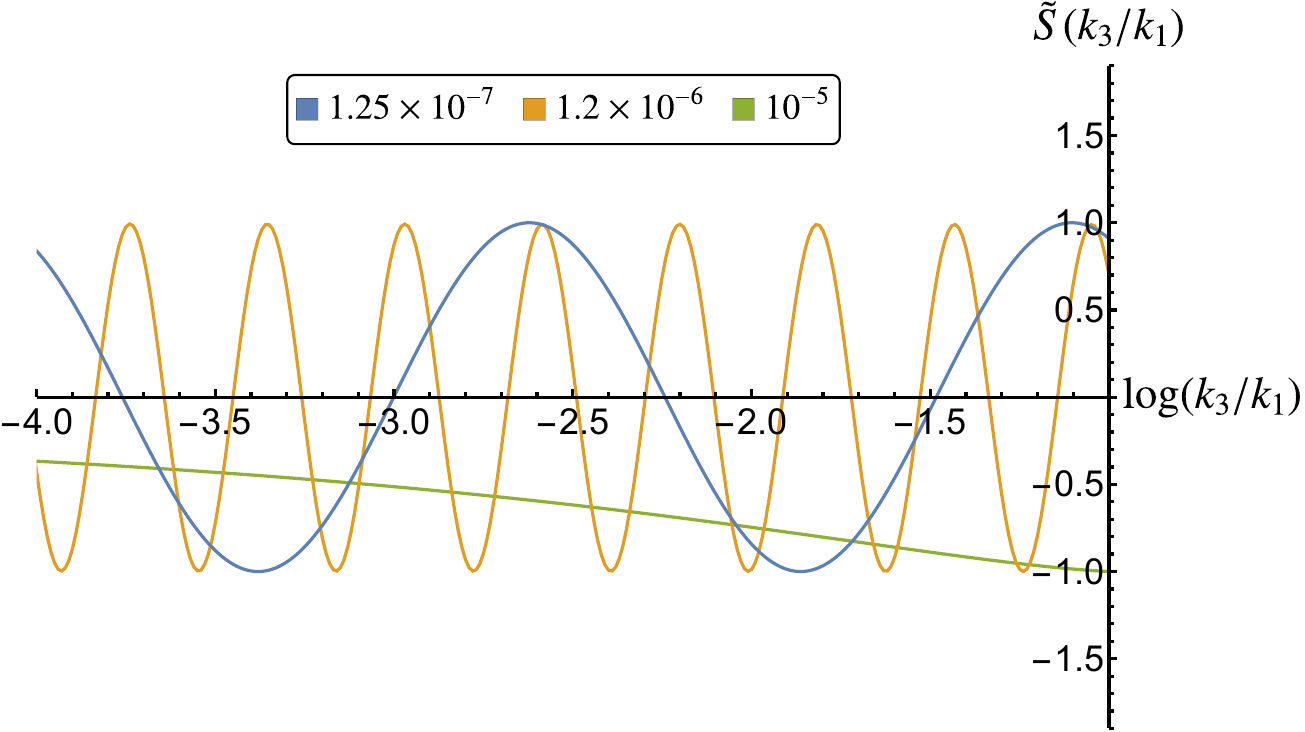}}
\caption{The spectral densities and $f_{NL}$ when the scalar bulk mass $m^2\sim 0$ for various values of UV brane mistunes, $\delta$. We also show some of the shape functions, which exhibit clear oscillatory behavior when there is a particle slightly above the critical mass, $3/2 H$.}
\label{fig:nu2}
\end{figure}

In Figure~\ref{fig:nu2} we show the spectral densities (left) and the values of $f_\text{NL}$ (right) for $\nu \approx 2$ for various values of $\delta = -2 \lambda$.  We see that the peak in the spectral density approaches the gap, and then dips below it as the value of the mistune, $\delta$, decreases.  There is a very sharp increase in the value of $f_\text{NL}$ when the particle passes across the gap, $3/2 H$.  The oscillatory features are prominent as the value of $\delta$ (and the position of the peak) increase, until eventually the continuum dominates the contributions to $f_\text{NL}$.  The shape functions (bottom) are shown for a few values of $\delta$ as well, with the oscillatory features showing up strongly when the particle feature is just above the gap.

\begin{center}
{\bf Near the BF Bound:  $m^2 \approx -4$}
\end{center}

On the CFT side, this choice of parameters corresponds to sourcing a double trace operator of the form
\beq
{\mathcal L} \ni \lambda {\mathcal O}^\dagger {\mathcal O},
\eeq
where the operator ${\mathcal O}$ has dimension close to 2, and at large $N$, the composite operator has dimension close to the sum.  The $\beta$ function corresponding to running of this operator is $\beta = - \lambda^2$~\cite{Witten:2001ua}.

On the 5D side, the bulk mass is taken close to the BF bound, so that $\nu \approx 0$.  The boundary mass is interpreted as the size of this coupling, and the AdS/CFT dictionary states that $m_0^2 = 2 (2-\lambda)$.  Since we have $\delta = m_0^2 - 2 (2-\nu)$, we thus read off $\delta = 2 (\nu-\lambda)$, and thus we can relate $\delta$ directly to the scaling dimension and the size of the deformation.

In order for this deformation to be nearly marginal, the UV brane localized mass is to be tuned against the bulk mass.  This is because the value of this mistune is associated with the coupling (and the $\beta$-function coefficient) associated with the nearly-marginal double trace operator that deforms the CFT and leads to a dilaton potential~\cite{Vecchi:2010em}.

It is very interesting that it is this precise range of parameters that can lead to an IR localized state that is near to the horizon, producing a ``cosmological quasiparticle" in the spectrum.  For the purposes of this paper, however the effect of this state on cosmological observables in the particular inflationary scenario we study is suppressed due to the small wave-function overlap between this IR localized state and the UV localized inflaton.  This is not expected to be the case in general, but is instead an artifact of the particular simple model of inflation that we have adopted.

It would be very interesting to explore models where the inflaton is \emph{also} IR localized, and dual to a composite state of the CFT.  In this case, we do not expect there to be a large wave-function overlap suppression factor, and there may be non-trivial cosmological collider signals of this horizon-bound particle.  This is beyond the scope of this current paper, but may be an exciting avenue for future work.

Even though there is no large particle contribution, the continuum contribution to the cosmological observables is still sizable.  At $\nu \approx 0$, and with a Hubble rate near it's currently allowed maximum, future experiments, particularly surveys of 21 cm emission, may be able to probe non-gaussianities arising from this operator.

\begin{figure}[h!]
\center{
\includegraphics[width=0.48\textwidth]{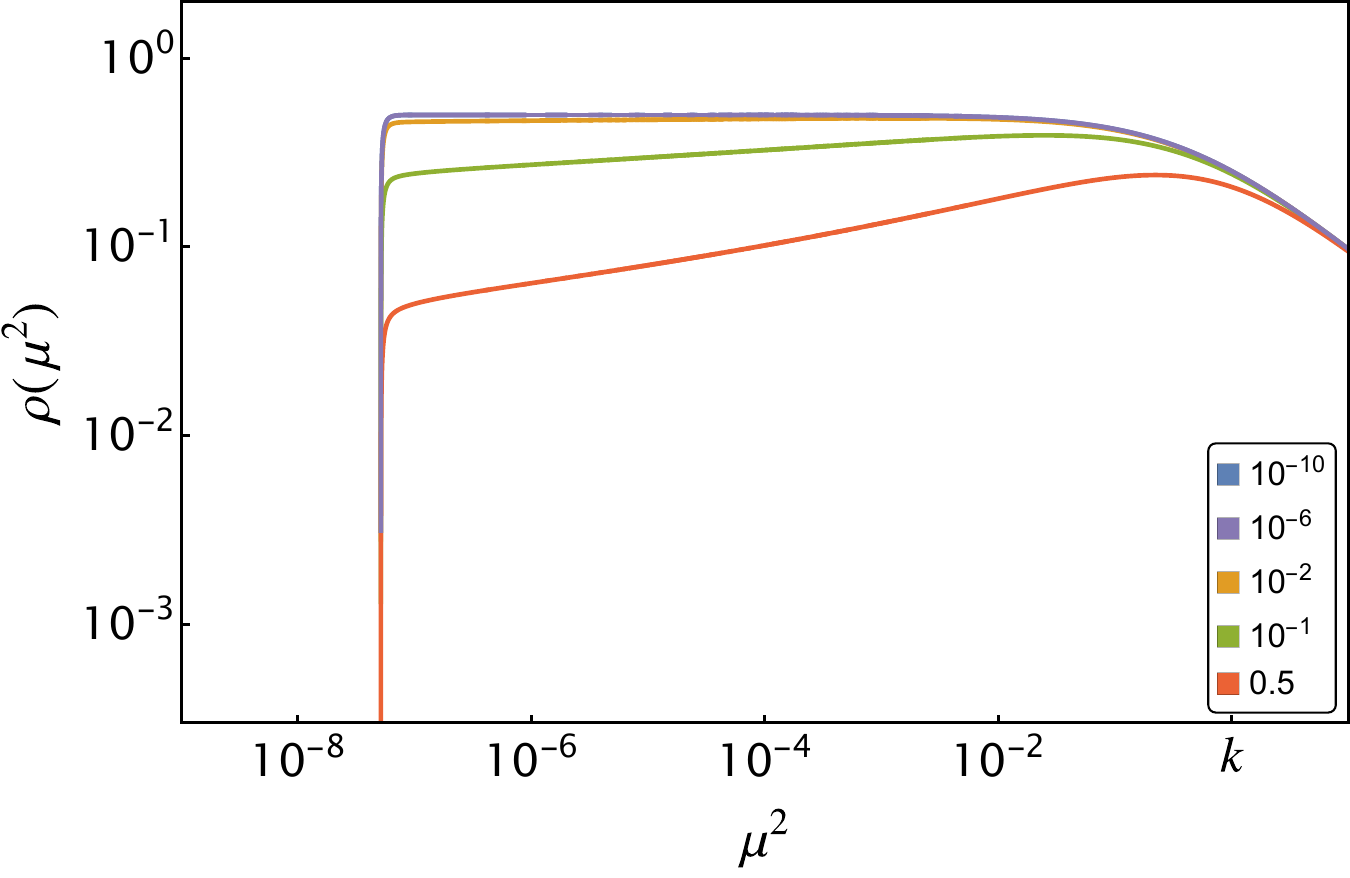}
\includegraphics[width=0.48\textwidth]{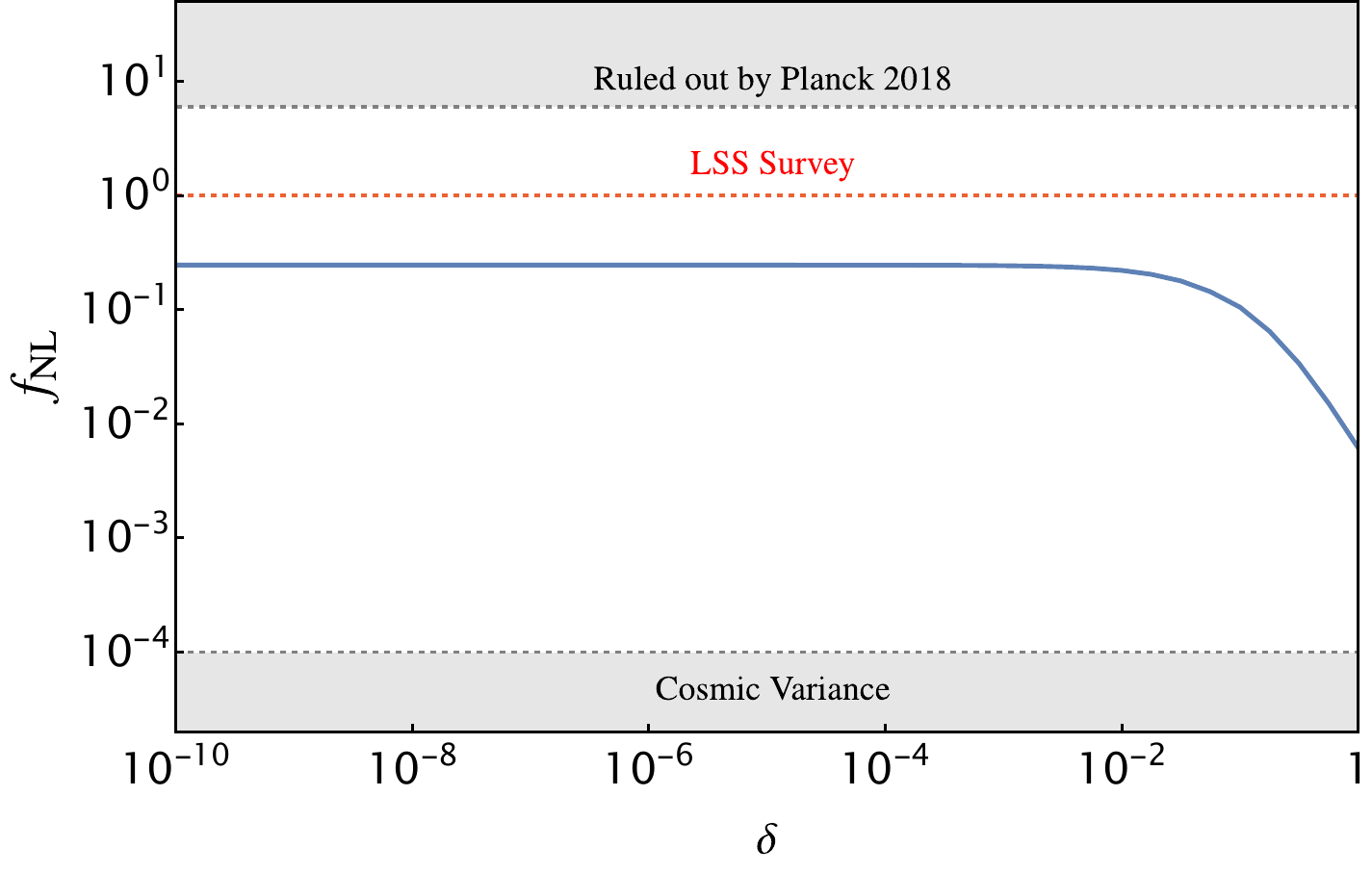}
\includegraphics[width=0.48\textwidth]{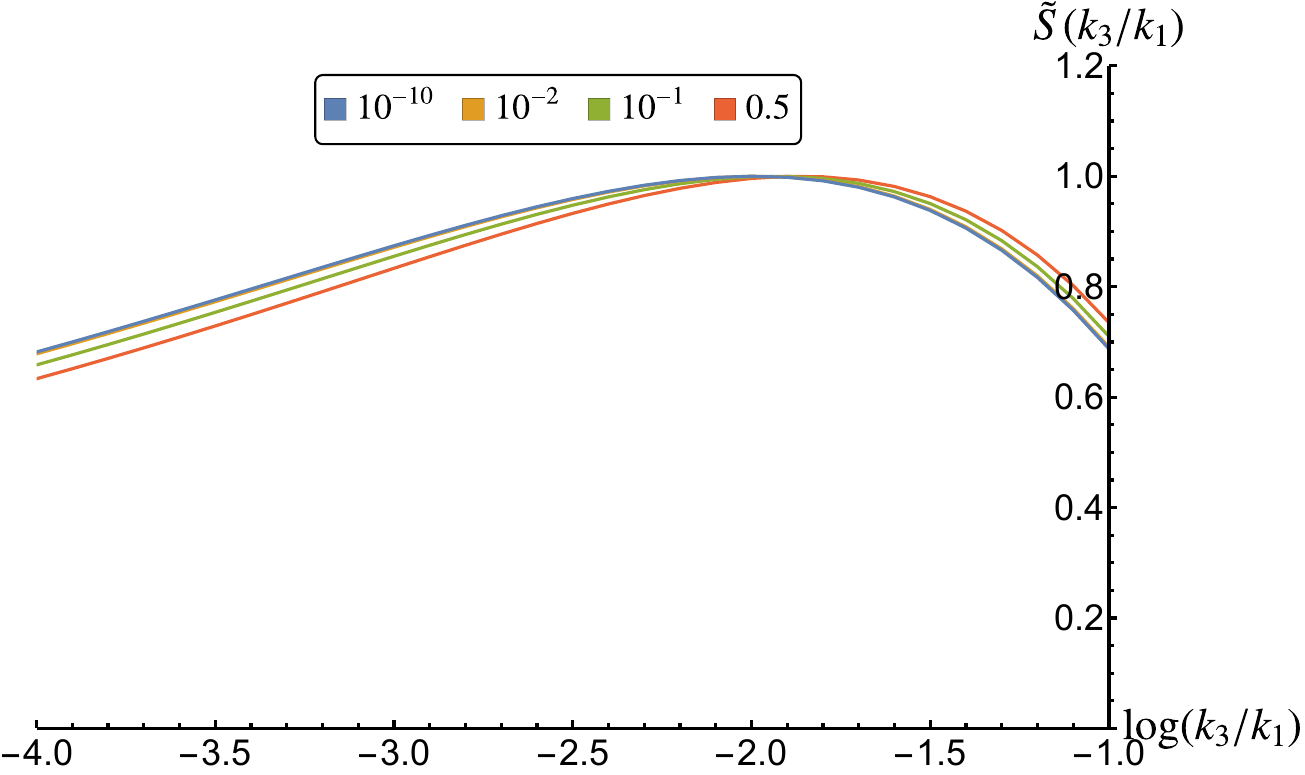}}
\caption{The spectral densities and $f_{NL}$ when the scalar bulk mass $m^2\sim -4$ for various values of UV brane mistunes, $\delta$.}
\label{fig:nu0}
\end{figure}

In Figure~\ref{fig:nu0} we show the spectral densities (left) and the values of $f_\text{NL}$ for $\nu = 0$, and for various values of $\delta = -2 \lambda$ (right).  We see that the shape of the spectral density quickly asymptotes to a fixed function as $\delta$ is decreased, which is consistent with the expectation that the modes are dominantly localized in the IR region of the geometry, near the horizon.  Only when $\delta$ is relatively large, giving a large effective localized mass to the field, does the spectral density begin to change shape.  This is due to the wave-function suppression from the large mass term.  This is reflected in the value of $f_\text{NL}$, which remains nearly constant until $\delta$ becomes sizeable, and in the shape functions, which are essentially the same for different values of $\delta$.

We emphasize that this is not necessarily the case for all models of inflation.  Near the BF bound, any particle-like features are localized in the IR, and thus couple extremely weakly to the UV localized inflaton.  We expect that in models where the inflaton is dual to a composite mode, and IR localized, that these results would be quite different.

Nevertheless, we see that the contributions to $f_\text{NL}$ are strong enough that future experiments may be able to probe this region of parameter space.

\section{Conclusions}

We have performed a study of the cosmological effects of holographic duals to strongly-coupled near-conformal dynamics that are weakly coupled to a fundamental inflationary sector.  Specifically, we explored Randall-Sundrum models with 5D scalar fields propagating in a space that is AdS up to deformations associated with cosmological expansion due to a rolling inflaton field that is localized on the UV cutoff brane.   The inflaton creates a UV brane tension that, at early times, is ``mistuned" against the bulk cosmological constant, leading to an early inflationary phase of the 5D universe where 4D slices of the geometry are approximately deSitter.

This deformation of the geometry creates a horizon in the infrared region, which cuts off the geometry, and can obscure the IR brane.  This is dual to the expansionary dynamics creating a deSitter temperature which exceeds the confinement scale of the dual strongly-coupled theory.  The resulting spectrum is that of a gapped continuum, with the gap being set by the deSitter temperature.  

This continuum begins precisely at the scale where modes contribute to non-analytic features in the bi-spectrum of scalar perturbations in the squeezed limit, and leads to a superposition of such contributions over a range of mass scales beginning at the gap, $\bar{\mu} = \frac{3}{2} H$.  Typically, this continuum does not contribute predominantly at any one scale, and so oscillatory features in the bi-spectrum are not always observable, though the shape function of the amplitudes follows a characteristic and universal $(k_3/k_1)^{3/2}$ scaling behavior.  

Particle like features \emph{can} however appear in the continuum due to mixing with a UV localized (fundamental) scalar mode.  The non-analytic contributions of these can rise above the continuum contributions, giving the ``smoking gun" oscillatory features in the shape function for $F_\text{NL}$.  There are significant regions of the parameter space that are allowed by current constraints, and that exceed the threshold associated with cosmic variance, and in some cases, there are contributions that have a unique shape due to the superposition of effects from a particle near the gap, and from the continuum.

We identified a novel kind of resonance dual to a composite mode with mass of order the Hubble scale, $H$ when the modes are IR localized (with bull mass $m^2 < -3$).  In 5D, these modes are localized near the horizon associated with the dS expansion.  While they do not contribute strongly to cosmological observables in the specific fundamental (or UV localized) inflationary model we employed, they may give significant contributions if the inflaton itself is part of the dual strongly coupled sector, or IR localized in the 5D description.  This is an exciting avenue for future work.

We also analyzed specific scenarios where the scalar field contributing to cosmological observables is associated with stabilization of the Planck-weak hierarchy in the present epoch.  The contributions to $f_\text{NL}$ are significant enough in these scenarios to be observable if the Hubble rate during inflation is not too far from its maximum observed value.

\section*{Acknowledgements}
We are grateful to Tom Hartmann, Ameen Ismail, Csaba Csaki, and Scott Watson for useful discussions during the preparation of this manuscript.  SJL is supported by the Samsung Science Technology Foundation under Project Number SSTF-BA2201-06. SJL is also supported by the National Research Foundation of Korea (NRF) funded
by the Korea government (MEST).  JH is thankful for continued hospitality at Cornell University throughout this work.
The work of JH, BS, and HL is supported in part by U.S. Department of Energy (DOE), Office of Science, Office of High 
Energy Physics, under Award Number DE-SC0009998.

\section*{Appendix A: The Normalized Spectral Density}
In this Appendix we derive the Green's function for the gapped scalar continuum in the inflating background, and associate with it a holographic spectral density (corresponding to the UV-UV brane propagator).  This spectral density appears in the 3-point inflationary correlators for the UV-brane localized inflaton.

The equation of motion in the coordinates defined by Eq.~\ref{eq:metansatz} is given by
\beq\label{eq:SLform}
	\Box_{\text{dS}}\Phi-\mathcal{O}_{z}\Phi=\lambda^2\Phi,
\eeq
where the first term on the LHS is the 4D deSitter Box operator, $\Box_{\text{dS}}\Phi = -\mu_4^2\Phi$. After the separation of variables, $\Phi=\sigma(\eta)\varphi(z)$, the second term defines a Sturm-Liouville problem in the $z$-coordinate $\mathcal{O}_z\varphi=-\mu^2\varphi$
\beq
	\mathcal{O}_z\varphi=\left(kz\right)^3\,G\partial_z\left(\frac{G}{(kz)^3}\varphi'\right)-\frac{m^2}{(kz)^2}\varphi.
\eeq
Then, Eq.~\ref{eq:SLform} gives an eigenvalue constraint equation
\beq
	-\mu_4^2+\mu^2=\lambda^2,
\eeq
where $\Phi$, $\sigma$, and $\varphi$ carry an implicit label of their corresponding eigenvalues, $\lambda^2$, $\mu_4^2$, and $\mu^2$. These functions satisfy the following orthogonality conditions 
\beq\label{eq:OrthoPhimu2}
	\int dz\,\varphi_{\mu^2}\varphi_{\mu'^2}\,w_z(z)=(2\pi)\delta(\mu^2-\mu'^2)\Theta\left(\frac{\mu^2}{H^2}-\frac{9}{4}\right)+\sum_{i}\delta_{\mu^2,\mu_i^2} \delta_{\mu'^2,\mu_i^2},
\eeq
\beq
	\int\, d\eta\,\sigma^{\dagger}_{\mu_4^2}(\eta)\sigma_{\mu_4'^2}\,w_4(\eta)=(2\pi)\delta(\mu_4^2-\mu_4'^2).
\eeq
Note that we separate out modes in the spectrum of solutions that reside in a continuum, and those which correspond to isolated discrete eigenvalues, $\mu_i$.

The corresponding completeness relations for the solutions are
\beq
	\left(\int_{\bar{\mu}^2}^{\infty}\frac{d\mu^2}{(2\pi)}+\sum_{\mu^2=\mu_i^2}\right)\varphi_{\mu^2}(z')\varphi_{\mu^2}(z)=\frac{1}{w_z(z)}\delta(z-z'),
\eeq
\beq
	\int_{-\infty}^{\infty}\frac{d\mu_4^2}{(2\pi)}\sigma^{\dagger}_{\mu_4^2}(\eta')\sigma_{\mu_4^2}(\eta)=\frac{1}{w_4(\eta)}\delta(\eta-\eta').
\eeq
Here, $w^{-1}_z(z)=(kz)^3\,G$ and $w^{-1}_4(\eta)=(H\eta)^4$ are the Sturm-Liouville weights.

We now have solutions to the equation of motion that allow us to express the Green's function as a notional ``sum" over modes:
\beq
	G^{(5)}=\text{``}\sum_{\lambda^2}\text{''} \frac{\Phi^{\dagger}\Phi}{\lambda^2}.
\eeq
More concretely, the Green's function is
\begin{multline}
	G^{(5)}(\eta,\eta',\vec{x}-\vec{x}',z,z')=\int\frac{d\mu_4^2}{(2\pi)}\int\frac{d^3k}{(2\pi)^3}e^{i\vec{k}\cdot(\vec{x}-\vec{x}')}\sigma^{\dagger}\left(-|\vec{k}|\eta'\right)\sigma\left(-|k|\eta\right)\\\times\left[\left(\int_{\bar{\mu}^2}^{\infty}\frac{d\mu^2}{(2\pi)}+\sum_{\mu^2=\mu_i^2}\right)\frac{\varphi^\dagger(z)\varphi(z')}{\lambda^2}\right].
\end{multline}
Note here that the spectrum is a continuum above the gap, $\bar{\mu}=3/2 H$, denoted by the integral in the square brackets, and a discrete set of modes below the gap, denoted by a sum in the square brackets over the masses of the discrete modes, $\mu_i$. We are interested in the UV brane to UV brane propagator, since our inflaton is taken to be localized there, which evaluates to:
\beq
	G^{(5)}|_{z=z'=z_0}=\left(\text{4D Integrals}\right)\times\left[\left(\int_{\bar{\mu}^2}^{\infty}\frac{d\mu^2}{(2\pi)}+\sum_{\mu^2=\mu_i^2}\right) \frac{|  \varphi(z_0) |^2}{\mu^2-\mu_4^2}\right].
\eeq
This takes the form of a two-point function in K\"ahl\'en-Lehmann representation where we interpret the term in brackets as an integral over a spectral density taking the form
\beq\label{eq:SpecDensityScalar}
	\rho(\mu^2)=\frac{1}{2\pi} \big\lvert\varphi_{\mu^2}(z_0)\big\rvert^2 \Theta(\mu^2-\bar{\mu}^2)+\sum_{i}\delta (\mu^2-\mu_i^2) \big\lvert\varphi_{\mu_i^2}(z_0)\big\rvert^2.
\eeq
The goal now is to find the fully normalized scalar field solutions in order to have a complete expression for this spectral density.  

The scalar solutions take the form
\beq\label{eq:ScalarSolwEig}
	\varphi_{\mu^2}=N_{\mu^2}\tilde{\varphi}_{\mu^2}=N_{\mu^2}\left(f^+_{\mu^2}+\alpha f^-_{\mu^2}\right),\,\text{with}\, N_{\mu^2}=N^C_{\mu^2}\Theta\left(\frac{\mu^2}{H^2}-\frac{9}{4}\right)+N^D_{\mu^2}\delta_{\mu^2,\mu^{*2}},
\eeq
where $\Theta$ is the Heaviside function that has support in the continuum ($\mu\geq \frac{3}{2}H$), and the kronecker delta corresponds to the eigenvalue of a light mode below the gap ($\mu=\mu^*<\frac{3}{2}H$).  The coefficient $\alpha$ is to be determined by imposing the UV brane boundary conditions $\frac{\varphi'}{\varphi}=\frac{1}{2\,(kz)G}m_0^2\bigg\vert_{z_0}$.  We take the $f_\pm$ solutions to be:
\beq
f_\pm (z) = \left( z H \right)^{2\pm \nu} ~_2 F_1 \left[ \frac{1}{4} \pm \frac{\nu}{2} - i \frac{\gamma}{2},\frac{1}{4} \pm \frac{\nu}{2} + i \frac{\gamma}{2}, 1 \pm \nu , - (z H)^2 \right].
\eeq

Using integration by parts, one can write down the orthogonality condition of the scalar solutions with different eigenvalues as
\beq
	\int_{z_0}^{\infty}\varphi_{\mu^2}\varphi_{\mu'^2}w_z(z)\,dz=\frac{1}{(\mu^2-\mu'^2)}\frac{G}{(kz)^3}\left[\varphi'_{\mu^2}\varphi_{\mu'^2}-\varphi_{\mu^2}\varphi'_{\mu'^2}\right]\bigg\rvert_{z_0}^{\infty},
\eeq
where $z=z_0$ is the location of the UV brane. For the moment we will allow the mass-term boundary conditions of $\varphi_{\mu^2}$ and $\varphi_{\mu'^2}$ to be different, associated with boundary mass terms $m_0^2$ and ${m_0'}^2$, respectively. We will take the limit where they coincide later. The UV brane piece with support in the continuum trivially vanishes in this limit. The asymptotics of the scalar solutions $f^{\pm}$ are
\beq
	f^{\pm}\to (z H)^{3/2 + i \gamma} \Sigma(\pm\nu,\gamma) + \text{h.c.}
\eeq
where $\nu=\sqrt{4+m^2}$, $\gamma=\sqrt{\frac{\mu^2}{H^2}-\frac{9}{4}}$, $\gamma'=\sqrt{\frac{\mu'^2}{H^2}-\frac{9}{4}}$. For modes below the gap, $\gamma$ is imaginary, and it is more convenient to use the tilde variables: $\tilde{\gamma}=\sqrt{9/4-(\mu/H)^2}$, $\tilde{\gamma}'=\sqrt{9/4-(\mu'/H)^2}$. We have defined
\beq
	\Sigma(\nu,\gamma)=\frac{2^{-\frac{1}{2}+\nu+i\gamma}}{\sqrt{\pi}}\frac{\Gamma(1+\nu)\Gamma(i\gamma)}{\Gamma\left(\frac{1}{2}+\nu+i\gamma\right)}.
\eeq
Note that below the gap, the only term that is asymptotically regular is proportional to $z^{-(\tilde{\gamma}+\tilde{\gamma}')}$, which then vanishes as we take $z\to\infty$. So, only the UV brane term contributes for particles below the gap. Above the gap, the UV boundary conditions eliminate the contribution evaluated at the UV brane.  There are many terms when the large z contribution are expanded, but only the ones that oscillate with wave number proportional to $\Delta\gamma\sim\gamma-\gamma'$ are important. This yields
\begin{align}\label{eq:OrthB4Delta}
\int_{z_0}^{\infty}dz\,\varphi_{\mu^2}\varphi_{\mu'^2}w_z(z)= & N^C_{\mu^2}N^C_{\mu'^2}\Theta\left(\frac{\mu^2}{H^2}-\frac{9}{4}\right)\Theta\left(\frac{\mu'^2}{H^2}-\frac{9}{4}\right) \nonumber \\ \times & \lim_{z\to\infty}\left[2 H^2 \text{Re}\left[F(\gamma,\gamma',\nu)\right]\frac{\sin\left((\gamma-\gamma')\log zH\right)}{(\gamma-\gamma')}\right] \nonumber \\
+ & \frac{1}{2} N^D_{\mu^2}N^D_{\mu'^2}\frac{1}{(kz)^4}\left(\lim_{\Delta m^2_0\to 0}\frac{\Delta m^2_0}{\Delta \mu^{2}}\right) \tilde{\varphi}_{\mu^2} \tilde{\varphi}_{\mu'^2}\bigg\vert_{z_0}  \delta_{\mu^2,\mu^{*2}}  \delta_{{\mu'}^2,\mu^{*2}},
\end{align}
where $\Delta\mu^{2}=\left(\mu^{2}-\mu'^{2}\right)$, and $\Delta m^2_0= m_0^2 - {m_0'}^2$ is the difference in the mass term boundary conditions between $\varphi_{\mu^2_z}$ and $\varphi_{\mu'^2_z}$. We have also defined
\beq
	F(\gamma,\gamma',\nu)=\left(\Sigma(\gamma,\nu)+\alpha\Sigma(\gamma,-\nu)\right)\left(\Sigma(-\gamma',\nu)+\alpha\Sigma(-\gamma',-\nu)\right).
\eeq
Next, we can replace the limit in the first term of Eq.\ref{eq:OrthB4Delta} $\lim_{z\to\infty}\frac{\sin\left((\gamma-\gamma')\log zH\right)}{\gamma-\gamma'}\to\pi\delta\left(\gamma-\gamma'\right)$, and $\lim_{\Delta m^2_0\to 0}\frac{\Delta m^2_0}{\Delta \mu^{2}} \rightarrow \frac{\partial m_0^2}{\partial \mu^2}$. 

Finally, we convert everything back to the dimensionful mass eigenvalues, $\mu^2$ and $\mu'^2$:
\begin{multline}
		\int_{z_0}^{\infty}dz\,\varphi_{\mu^2}\varphi_{\mu'^2}w_z(z)=2\pi \delta\left(\mu^2-\mu'^2\right)\Theta\left(\frac{\mu^2}{H^2}-\frac{9}{4}\right)  \left| N^C_{\mu^2} \right|^2 2 H^4 \text{Re}\left[F(\mu^2,\mu'^2,\nu)\right]\sqrt{\frac{\mu^2}{H^2}-\frac{9}{4}} \\
+\delta_{\mu^2,\mu^{*2}}  \delta_{{\mu'}^2,\mu^{*2}} | N^D_{\mu^{2}} |^2 \frac{1}{2 (kz_0)^4}  \frac{\partial m_0^2}{\partial \mu^2} \left| \tilde{\varphi}_{\mu^{2}} (z_0)\right|^2.
\end{multline}
We can now read off the normalizations by comparing this with Eqs.~\ref{eq:OrthoPhimu2} and~\ref{eq:ScalarSolwEig}:
\beq
	\big\lvert N^C_{\mu^2}\big\rvert^2=\frac{1}{2 H^4}\frac{1}{\sqrt{\frac{\mu^2}{H^2}-\frac{9}{4}}}\frac{1}{\text{Re} \left[F(\mu^2,\nu)\right]},~~~~\big\lvert N^D_{\mu^2}\big\rvert^2=\frac{2 (kz_0)^4}{\frac{\partial m_0^2}{\partial \mu^2}} \frac{1}{\big\lvert \tilde{\varphi}_{\mu^2}(z_0)\big\rvert^2}.
\eeq
With this information, we can evaluate Eq.~\ref{eq:SpecDensityScalar} and obtain the spectral density:
\beq
	\rho\left(\mu^2_z\right)=\left[\frac{1}{4\pi H^4} \frac{1}{\sqrt{\frac{\mu^2}{H^2}-\frac{9}{4}}}\frac{\big\lvert\tilde{\varphi}_{\mu^2}(z_0)\big\rvert^2}{\text{Re}\left[F(\mu^2,\nu)\right]}\right]\Theta\left(\frac{\mu^2}{H^2}-\frac{9}{4}\right)+ \frac{2 (k z_0)^4}{\frac{\partial m_0^2}{\partial \mu^2}}\delta\left(\mu^2-\mu^{*2}\right).
\eeq

In the case that there is a particle below the threshold for the continuum and $\nu > 1$ (corresponding to UV brane localization), the relative normalization of the particle contribution to the spectral density can be evaluated analytically using Eq.~\ref{eq:mu2approx} to give 
\beq
	\frac{2 (k z_0)^4}{\frac{\partial m_0^2}{\partial \mu^2}} \approx 2 (kz_0)^4 (\nu-1).
\eeq

%%%%%%%%%%%%%%%%%%%%%%%%%%%%%%%%%%%%%%%%%%%%%%%%%%%%%%
%%%%%%%%%%%%%%%%%%%%%%%%%%%%%%%%%%%%%%%%%%%%%%%%%%%%%%

%%%%%%%%%%%%%%%%%%%%%%%%%%%%%%%%%%%%%%%%%%%%%%%%%%%%%%
%%%%%%%%%%%%%%%%%%%%%%%%%%%%%%%%%%%%%%%%%%%%%%%%%%%%%%

\end{document}